\documentclass[a4paper,11pt]{article}
\pdfoutput=1 

\usepackage{jcappub} 

\usepackage[T1]{fontenc} 

\usepackage[english]{babel}
\usepackage[utf8]{inputenc}
\usepackage{graphicx}
\usepackage{slashed}
\usepackage{color}
\usepackage[dvipsnames]{xcolor}
\usepackage{rotating}
\usepackage{ulem}
 \usepackage{float}
 \usepackage{blindtext}
 \usepackage{multirow}

\newcommand{\be}{\begin{equation}}
\newcommand{\ee}{\end{equation}}
\newcommand{\bea}{\begin{eqnarray}}
\newcommand{\eea}{\end{eqnarray}}

\graphicspath{{fig/}{./figures/}} 

\numberwithin{equation}{section}

\title{Constraining self-interacting dark matter with scaling laws of observed halo surface densities}

\author[a]{Kyrylo Bondarenko,}
\author[a]{Alexey Boyarsky,}
\author[b]{Torsten Bringmann}
\author[b]{and Anastasia Sokolenko}

\affiliation[a]{Intituut-Lorentz, Leiden University, Niels Bohrweg 2, 2333 CA Leiden, The Netherlands}
\affiliation[b]{Department of Physics, University of Oslo,Box 1048, NO-0371 Oslo, Norway}

\emailAdd{bondarenko@lorentz.leidenuniv.nl}
\emailAdd{boyarsky@lorentz.leidenuniv.nl}
\emailAdd{torsten.bringmann@fys.uio.no}
\emailAdd{anastasia.sokolenko@fys.uio.no}

\abstract{ 
The observed surface densities of dark matter halos are known to follow a simple 
scaling law, ranging from dwarf galaxies to galaxy clusters, with a weak dependence
on their virial mass. Here we point out that this can not only be used to 
provide a method to determine the standard relation between halo mass and concentration,
but also to use large samples of objects in order to place constraints on dark matter self-interactions 
that can be more robust than constraints derived from individual objects. 
We demonstrate our method by considering a sample of about 50 objects distributed across
the whole halo mass range, and by modelling the effect of self-interactions in a way similar to what 
has been previously done in the literature. Using additional input from simulations
then results in a constraint on the self-interaction cross 
section per unit dark matter mass of about $\sigma/m_\chi\lesssim 0.3$\,cm$^2$/g. 
We expect that these constraints can be significantly improved in the future, and made more robust,
by {\it i)} an improved modelling of the effect of self-interactions, both theoretical and
by comparison with simulations,
{\it ii)} taking into account a larger sample of objects and {\it iii)} by reducing
the currently still relatively large uncertainties that we conservatively assign to the surface 
densities of individual objects. The latter can be achieved in particular by using kinematic 
observations to directly constrain the average halo mass inside a given radius, 
rather than fitting the data to a pre-selected profile and then reconstruct the mass. For a velocity-independent cross-section, our current result is 
formally already somewhat smaller than the range $0.5-5$\,cm$^2$/g that has been invoked to 
explain potential inconsistencies between small-scale observations and expectations in the standard 
collisionless cold dark matter paradigm.}

\begin{document}
\maketitle
\flushbottom

\section{Introduction}
\label{sec:intro}

On cosmological scales, dark matter (DM) is about five times as prevalent as 
ordinary matter \cite{Ade:2015xua}, and known to be the main driver of structure 
formation. The paradigm of cold, collisionless dark matter (CDM), one of the main ingredients of the
cosmological concordance model, has been remarkably successful in describing the observed
distribution and properties of structures in the universe \cite{Springel:2005mi,Vogelsberger:2014kha}.
In apparent contradiction to this success, however, current observations do not actually constrain the DM 
self-interaction cross section to be smaller than that of the strong interaction between nucleons (for a 
recent review, see \cite{Tulin:2017ara}), which is many orders of magnitude less stringent 
than corresponding bounds on DM interacting with standard model particles \cite{Aprile:2017iyp,Cui:2017nnn}. 
Self-interacting DM (SIDM) thus remains a fascinating option which, if directly confirmed 
observationally, would significantly reduce the number of possible DM candidates from particle
physics. Such observations would, furthermore, offer a window into the particle properties of DM 
that may be impossible to access by other means -- a fact which has created significant attention in recent years 
(see, e.g.~Refs.~\cite{Bringmann:2013vra, Kaplinghat:2015aga,Valli:2017ktb}).

The typical phenomenological handle on SIDM models, and in fact the context in which the very
idea of SIDM was proposed in the first place \cite{Spergel:1999mh}, are observables related to structure 
formation at galactic scales and below. In particular, it has been demonstrated 
\cite{Loeb:2010gj,Vogelsberger:2012ku,Peter:2012jh,Zavala:2012us,Aarssen:2012fx,Elbert:2014bma,Kamada:2016euw,Robertson:2017mgj}
that SIDM could alleviate all of the potential  small-scale problems of $\Lambda$CDM cosmology 
\cite{Bullock:2017xww}, most notably the ``core-cusp'' \cite{Flores:1994gz,1994Natur.370..629M}, ``too-big-to-fail''
\cite{2011MNRAS.415L..40B,BoylanKolchin:2011dk}, ``diversity''  \cite{Oman:2015xda,Oman:2016zjn}
and (for late kinetic decoupling) ``missing satellites'' problems \cite{Moore:1999nt,Klypin:1999uc,Fattahi:2016nld}.
These solutions require a DM self-scattering cross section per unit DM mass of the order of 
$\sigma/m_\chi\sim1$\,cm$^2$/g, close to current exclusion limits.
But even if all current small-scale discrepancies between $\Lambda$CDM observations and expectations are resolved, in the sense that they can be attributed to  baryonic effects, the  uncertainty in Milky Way mass, cosmic variance and observational uncertainties (see e.g.~\cite{Mashchenko:2007jp,2012MNRAS.424.2715W,2015MNRAS.454..550G,Oman:2016zjn}), SIDM remains an intriguing possibility -- not the least given the absence of any undisputed positive results in searches for DM particles. 
Current constraints derive from a large number of
observations on different scales, see again \cite{Tulin:2017ara} for an overview, but their
common feature is that they are typically obtained from individual objects. Given the
modelling uncertainties of the respective objects, ranging from their individual formation history to
their baryonic content and its potential interference with the effects of SIDM, this is not unproblematic.

Here, we introduce a new way to constrain DM self-interactions that instead relies on 
{\it ensembles} of many astrophysical objects, thereby reducing the systematic uncertainties
related to individual objects. Concretely, we revisit the well-known observed scaling relation between
surface density and halo mass \cite{Donato:2009ab,Gentile:2009bw,Boyarsky:2009rb,Boyarsky:2009af}, which
is essentially understood in $\Lambda$CDM cosmology as a reflection of a similar relation 
between halo mass and concentration. 
We investigate how this relation is affected by the central cores  observed in the DM distributions 
in various (dwarf) galaxies and (possibly) also galaxy clusters. Assuming that these cores can 
exclusively be explained in terms of SIDM, 
we derive an experimentally robust upper bound on the effect of
DM self-interactions once we take into account additional constraints deriving from a direct 
comparison to SIDM simulations.
The final translation of this bound to a constraint on the physical self-interaction cross section per 
unit mass, $\sigma/m_\chi$, is necessarily somewhat less robust as it involves less certain
theoretical modelling of the effect of DM self-interactions.
Even when taking this into account, we arrive at an upper bound on $\sigma/m_\chi$
that is 
competitive compared to existing bounds in the literature. We note again
that this is mainly the result of the relatively large number of objects that we include in the
analysis, combined with crucial input from simulations about quantities that cannot be directly 
constrained observationally.

This article is organized as follows. We start in Section \ref{sec:cores} by reflecting on how 
DM self-interactions would change the halo density profiles expected in $\Lambda$CDM 
cosmology. In Section \ref{sec:sd}, we then introduce the observed surface density
scaling relation, as well as its  $\Lambda$CDM explanation, and derive how the surface
density is expected to change as a function of the DM self-interaction cross section. 
We finally derive constraints on the latter in Section \ref{sec:stat}, and discuss them,
before concluding in Section \ref{sec:conc}. In two Appendices, we describe in more
detail the halo objects that we include in our analysis (App.~\ref{app:data}) and 
briefly review how the halo age depends on its mass in $\Lambda$CDM 
cosmology (App.~\ref{app:age}).

\section{Halo profiles for self-interacting dark matter}
\label{sec:cores}

Numerical simulations of gravitational clustering in $\Lambda$CDM cosmology reveal 
that collision-less DM halos have a universal density profile that, at all redshifts and masses, 
roughly follows the form suggested by Navarro, Frenk and White (NFW) 
\cite{Navarro:1995iw,Navarro:1996gj}:
\be
    \rho_{\text{NFW}}(r) = 
    \frac{\rho_s}{(r/r_s)^\gamma (1 + (r/r_s))^{3-\gamma}}.
    \label{eq:NFW}
\ee
Here, the scale radius $r_s$ marks the position where the profile has a slope of
$d(\log\rho)/d(\log r)=-(3+\gamma)/2$, and hence  the transition between the slopes encountered
in the inner and the outer part of the halo. At redshift zero, CDM halos have mostly converged to a
negative
inner slope of $\gamma\approx 1$, with some scatter, and this is what we will use in 
the following when referring to the NFW profile. More recent simulations sometimes
tend to prefer an Einasto profile \cite{1965TrAlm...5...87E}, which is slightly shallower in the
very central parts of the halo, but this difference will not affect our discussion.

It should be stressed that the above result only holds for DM-only simulations. Including
baryons can both lead to significantly steeper, even cuspier DM profiles -- mostly due to a process known 
as adiabatic contraction \cite{Blumenthal:1985qy,Gnedin:2004cx,Gustafsson:2006gr} -- and 
to the formation of much shallower profiles in the inner parts, often referred to as {\it cores}, due to feedback 
from star formation and supernovae \cite{Sawala:2015cdf}. Despite the enormous numerical 
challenges, there has recently been significant progress in including baryonic effects in 
hydrodynamical simulations of structure formation even at cosmological 
scales \cite{Schaye:2014tpa, Schaller:2014uwa, Wang:2015jpa, Tollet:2015gqa}. However, 
this comes at the price of implementing phenomenological prescriptions, rather than
prescriptions based on first principles, that have to be tuned to match one class of observations
in order to successfully ``predict'' another.
In this sense, a complete understanding of all relevant scales and processes  is still missing.
We will try to avoid these difficulties by mostly focussing on systems where the effect
of baryons is subdominant (but get back to this discussion in Section \ref{sec:disc}).

The main effect of DM self-interactions, as we will see, is to isotropize the DM phase-space distribution $f$.
In other words, the claim is not only a Maxwellian velocity distribution,
\be
    f(\mathbf{v},\mathbf{r}) = N(\mathbf{r}) \exp\left(- \dfrac{\mathbf{v}^2}{2\sigma_v^2(\mathbf{r})}\right),
    \label{eq:eq}
\ee
but that the velocity dispersion $\sigma_v$ is (approximately) 
constant at least in some region of the halo, namely in its central parts where the DM density and
hence the collision rate is highest. This behaviour has been confirmed by numerical simulations
\cite{Vogelsberger:2012ku, Rocha:2012jg, Elbert:2014bma} for cross sections in the ballpark that we are interested in, 
$\sigma/m_\chi\sim1$\,cm$^2$/g, 
showing results broadly in agreement with the expectations originally formulated along with 
the SIDM proposal \cite{Spergel:1999mh}. It is worth noting that a constant $\sigma_v$ is only
possible in the weakly interacting regime for the DM particles. The entropy increase in 
the inner parts then leads to a formation of an inner  core of roughly constant density  
-- again confirmed by numerical simulations.
 For much larger cross sections, on the other hand,
DM would behave as a collisional gas, leading to an isothermal density profile with $\rho\propto r^{-2}$ 
(see, e.g., \cite{1987gady.book.....B,Pollack:2014rja}).

We can easily check that we are indeed in the weakly interacting regime, 
motivating the claim of a constant $\sigma_v$, by looking at the mean free path 
$\lambda$ of the DM particles,
\be
\lambda \equiv \frac{1}{\sigma n_\chi}
\simeq 4.8 \text{ kpc } \left( \frac{1 \text{ cm}^2/\text{g}}{\sigma/m_\chi} \right)
\left( \frac{1 M_{\odot}/\text{pc}^3}{\rho_\chi} \right).
\label{eq:meanfreepath}
\ee
This is clearly larger than the sub-kpc cores reported in dwarf galaxies, for realistic 
core densities $\rho_{\rm core} = \mathcal{O}(1)~M_{\odot}/\text{pc}^3$ for dwarf galaxies~\cite{Walker:2009zp} and $\rho_{\rm core} = \mathcal{O}(0.1)~M_{\odot}/\text{pc}^3$ for clusters~\cite{Newman:2012nw}. For these cross sections, a DM particle thus typically only scatters
at most a few times during the whole halo lifetime $t_{\rm age}$ even though it may pass through the core 
region much more often. In the outer parts of the halo, on the other hand, scattering is so rare that the
standard (NFW) DM profile should be unaffected.
When estimating the region of equilibrium $r<r_{\rm SIDM}$, for which we have
$\sigma_v\sim const.$, we should thus refer to the {\it average}
density inside this region,  $\langle\rho_\chi\rangle_{\rm SIDM}$, to compute the mean free path. We can 
do so by demanding that the average time
between collisions should roughly equal the halo time, i.e.\footnote{
This is essentially the same ansatz as in Ref.~\cite{Kaplinghat:2015aga}, with the most important
difference that we use $\langle\rho_\chi\rangle$ rather than $\rho(r_{\rm SIDM})$.
}
\be
 \frac{\lambda}{v_\chi}=
 \frac{m_\chi}{\sigma v_\chi \langle\rho_\chi\rangle_{\rm SIDM}}=t_{\rm age}/\xi \,.
 \label{sidm_ansatz}
\ee
Here, $v_\chi=4/\sqrt{\pi}\, \sigma_v$ is the average {\it relative} 
velocity of the DM particles 
inside the core,
and we have  allowed for an unknown factor $\xi\gtrsim1$ to account 
for the fact that, for scattering between 
two non-relativistic particles, the relaxation time (the time-scale needed to achieve thermal equilibrium) is 
indeed somewhat larger than the scattering rate (while for scattering between a DM particle and 
a relativistic particle of energy $E\ll m_\chi$, the difference would be $\xi\sim \sqrt{m_\chi/E}$
and hence {\it much} larger \cite{Hofmann:2001bi}). 

In order to relate $v_\chi$ to the circular velocity $v_c$ let us consider the Jeans equation, 
\begin{equation}
    \frac{\partial (\sigma_v^2 \rho_\chi)}{\partial r} = -\rho_\chi 
    \frac{\partial \Phi}{\partial r},
    \label{eq:Jeans}
\end{equation}
where $\Phi$ is the gravitational potential and it has been used that SIDM halos should be isotropic
and spherically symmetric in their inner parts \cite{Spergel:1999mh,Vogelsberger:2012ku}. 
For the range of radii where $\sigma_v$ is constant, $r<r_{\rm SIDM}$, we can rewrite the Jeans 
equation in the simple form
\be
  \sigma_v^2={\alpha^{-1}}\,v_c^2={\alpha^{-1}(r)}\,\frac{GM(r)}{r}\,,
  \label{eq:vcrel}
\ee
where
\be
\alpha = -\frac{r}{\rho_\chi}\frac{d\rho_\chi}{dr}
\label{eq:alphadef}
\ee
is the logarithmic slope of the DM profile at radius $r$, and $M(r)$ denotes the halo mass inside $r$.
This leads, finally, to the following implicit definition of $r_{\rm SIDM}$:
\be
\langle\rho_\chi\rangle^3_{\rm SIDM} r_{\rm SIDM}^2\equiv\left(\frac{3}{4\pi}\right)^3 M^3(r_{\rm SIDM})\, r_{\rm SIDM}^{-7}
= \frac{3}{64 G} \frac{\xi^2\alpha_{\rm SIDM}}{(\sigma/m_\chi)^2 t_{\rm age}^2}
\,.
 \label{rsidmimp}
\ee
Here, $\alpha_{\rm SIDM}$ is the (negative) logarithmic slope of the SIDM profile at $r=r_{\rm SIDM}$;
as we will see below, its value falls into the range $1\lesssim\alpha_{\rm SIDM}\leq2$. 
For $t_\mathrm{age}$ in the above expression, we will use the average halo age as a function 
of the virial mass adopted from 
Refs.~\cite{Ludlow:2013vxa,Ludlow:2016ifl,Liu:2012kd}, see Appendix \ref{app:age} for further details.

\bigskip
Before continuing, let us comment on why it is so challenging to directly relate the scale 
$r_{\rm SIDM}$ to the observed core size, $r_{\rm core}\leq r_{\rm SIDM}$, which would 
provide a handle on the scattering cross section that is observationally easy to access.
It has been argued \cite{Kaplinghat:2015aga,Kamada:2016euw,Elbert:2016dbb,Creasey:2016jaq,Valli:2017ktb} 
that this 
question can be fully resolved by considering the Jeans equation. 
In particular, when assuming $\sigma_v(r)\sim const.$ and using Poisson's equation,
as well as neglecting the contribution of baryons to the gravitational potential, Eq.~(\ref{eq:Jeans}) simplifies to
\begin{equation}
    \frac1{r^2}\frac{\partial}{\partial r}\left(\frac{r^2}{\rho_\chi}\frac{\partial \rho_\chi}{\partial r}\right) \approx -\frac{4 \pi G}{\sigma_v^2}\rho_\chi\,.
    \label{eq:Jeanssimp}
\end{equation}
It can be easily checked that the isothermal sphere, $\rho(r)=\sigma_v^2/(2\pi Gr^2)$, is a solution to this equation;
as discussed above, this is the physical solution for cross sections {much} larger than what we are interested in here.
There is, however, also another class of solutions, with $\rho'_\chi(0)=0$,
which describe a roughly constant density at 
small radii, and a profile approaching the isothermal sphere solution for large radii. The phenomenological 
{\it modified} \mbox{(or pseudo-)} isothermal sphere,
\begin{equation}
    \rho_{\text{ISO}}(r) = 
    \frac{\rho_0}{ 1 + r^2/r_0^2}\,,
    \label{eq:ISO}
\end{equation}
is a very good approximation to this class of solutions, at least for $r\ll r_0$ and $r\gg r_0$. 
Further {\it phenomenological} profiles that are commonly used to describe cored density distributions, and fit 
observational data, are the cored NFW and Burkert \cite{Burkert:1995yz} profiles,
see Appendix \ref{app:data},
which reproduce the expected scaling of the NFW profile rather than that of the isothermal sphere at large radii -- 
where in fact we cannot expect a constant $\sigma_v$ anymore.

\begin{figure}[t]
  \centering
   \includegraphics[width=0.48\textwidth]{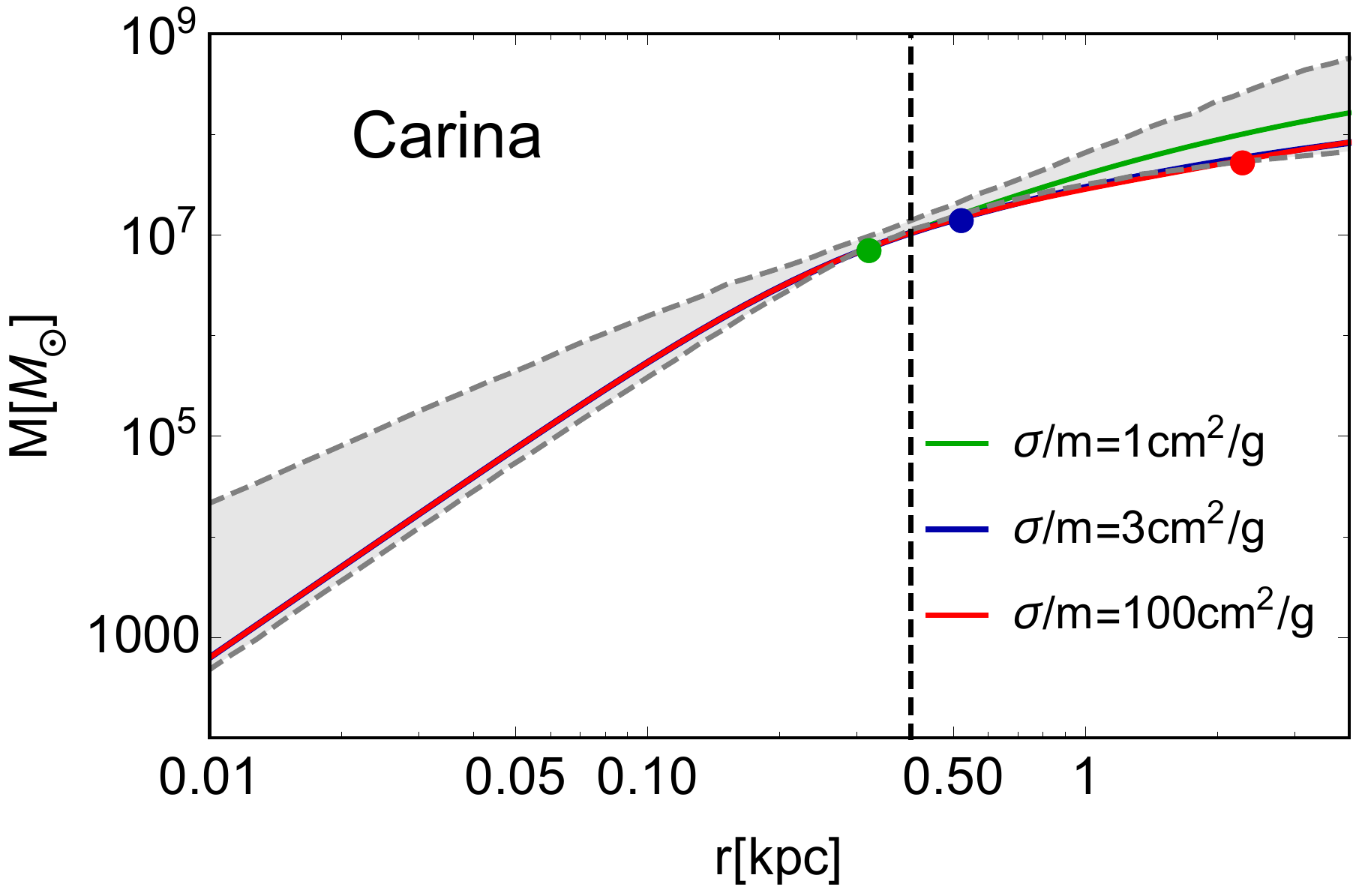}
   ~\includegraphics[width=0.48\textwidth]{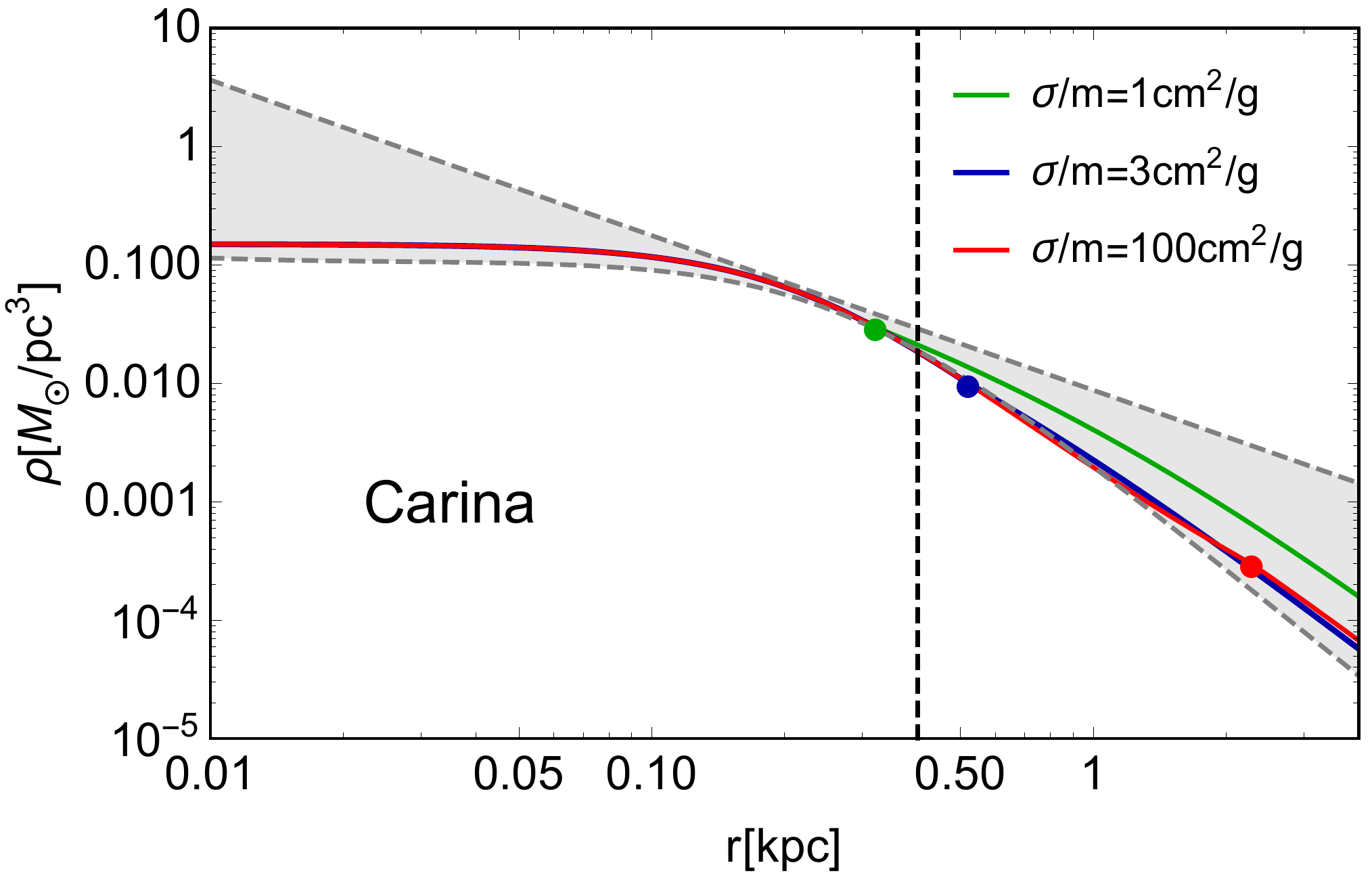}
   \caption{
  The mass and density profiles of the dwarf spheroidal galaxy Carina. The shaded regions are reconstructed values 
  from kinematic data measured within 1\,kpc radius for this object~\cite{walker_private}. The green, blue and red 
  lines are examples of SIDM profiles adopting the method from Ref.~\cite{Kaplinghat:2015aga}, with the
  slight modification given in Eq.~(\ref{sidm_ansatz}). for an increasing
  value of the self-interaction cross section as indicated. The dots represent the radius $r_{\text{SIDM}}$ in each
  of these cases, and the vertical dashed line is the half-light radius.}
   \label{fig:Carina}
\end{figure}

The cored solutions to Eq.~(\ref{eq:Jeanssimp}) still have one free parameter, for a given $\sigma_v$,
which can equivalently be chosen as the central density or core size (for the Burkert parametrization, e.g.,
the core size is given by $r_0=\sigma_v/\sqrt{2\pi G \rho_0}$). 
The issue is that the relation between the core size (or core density) and $r_{\rm SIDM}$ 
very strongly depends on the assumed {\it matching conditions} to the asymptotic NFW profile 
--
typically taken to be that self-interactions do not change the mass inside any radius larger
than $r_{\rm SIDM}$, and that the density profile remains continuous at $r=r_{\rm SIDM}$
 -- 
and hence on the part of the profile that should not be affected by DM self-interactions. We illustrate 
this in Fig.~\ref{fig:Carina} for the case of the dwarf spheroidal galaxy Carina, 
where the grey area indicates the range of the mass profile consistent with dynamic observations of the stellar 
population \cite{walker_private}. We note that this is a somewhat extreme example, picked for the
sake of our argument, where the whole range of kinematic observations is compatible with 
a cored density profile. Choosing one central density and core size consistent with these data, we plot the 
resulting profile for a large range of self-interaction cross sections $\sigma/m_\chi$.  Clearly, 
dynamical observations of this object hardly constrain $r_{\rm SIDM}$, allowing $r_{\rm SIDM}\gg r_c$ 
even if the core size $r_c$ is well measured 
 (we obtain qualitatively the same result if we replace in Eq.~(\ref{sidm_ansatz}) 
 $\langle\rho_\chi\rangle\to\rho_\chi$, as in Ref.~\cite{Kaplinghat:2015aga}).
A much more stringent constraint on the scale of self-interactions instead results 
from requiring that the asymptotic behaviour of the NFW profile matches the average 
expectation of $\Lambda$CDM cosmology \cite{Kaplinghat:2015aga,Valli:2017ktb}. 
As already indicated, classical dwarf spheroidal galaxies like Carina are the most extreme objects in this
respect. For other objects -- like the dwarf galaxies considered in \cite{Kaplinghat:2015aga}, which are
not Milky Way satellites -- the outer (NFW) part of the profile is typically much better constrained even without
taking  into account cosmological priors. 
Still, there is a remaining worry that the inferred (bound on the) cross section depends to 
some extent on the choice of matching conditions with the asymptotic NFW profile rather than only 
on the (not directly observable) physical size of the self-interaction region $r\lesssim r_{\rm SIDM}$.

In conclusion, there is a considerable theory uncertainty in how to relate, from first principles, the observationally 
accessible core size of individual halos (as opposed to the theoretically well-defined, but observationally inaccessible 
scale of self-interactions) to the self-interaction cross section. Concretely, the method
typically adopted so far may allow larger ratios between  $r_{\rm SIDM}$ and core size than what one
would expect from the underlying physics. In fact, it is quite conceivable that the 
exact core size even depends to some degree on the formation history and hence on environmental effects.
In this article we will instead fix a related quantity, namely the ratio between average densities inside the 
core and $r_{\rm SIDM}$, by directly comparing it to results taken from {\it simulations}.
Technically speaking, from the point of view of the Jeans analysis, this amounts to adopting a different set 
of boundary conditions to solve Eq.~(\ref{eq:Jeans}). 
Furthermore, we will make use of observed scaling relations in {\it ensembles} of astrophysical 
objects which are more robust to astrophysical uncertainties than the analysis of individual objects.

\section{Halo surface densities}
\label{sec:sd}

\subsection{Observations}

Let us define the {\it mean surface density}, sometimes also referred to as Newtonian acceleration, 
of a halo as
\be
\Sigma (r) \equiv \frac{M(r)}{\frac{4}{3} \pi r^2} \equiv \langle \rho \rangle_r r,
\label{def:sigma}
\ee
where
\be
 M(r)=4\pi \int_0^r {r'}^2 \rho(r') dr'
\ee
is the mass enclosed inside the radius $r$ (we assume that this mass is dominated by the DM component).
For a cored profile, this quantity is maximized at a radius close to the core radius.\footnote{For the modified 
isothermal sphere, e.g., we have $\Sigma_{\rm max}=\Sigma(1.515\,r_0)=1.07\,\Sigma(r_0)$. For the Burkert profile, the 
maximal value is obtained even closer to the core radius: 
$\Sigma_{\rm max}=\Sigma(0.96\,r_0)=1.001\,\Sigma(r_0)$.
}
For the NFW profile $\Sigma(r)$ is approximately constant for $r\ll r_s$ and reaches its maximum for $r\to0$, 
with $\Sigma(0)=1.15\,\Sigma(0.1\,r_s)=2.62\,\Sigma(r_s)$.
We will denote this maximal value as $\Sigma_{\rm max}$.
Refs.~\cite{Donato:2009ab,Gentile:2009bw} argued that for galaxies (from dSph to ellipticals) $\Sigma_{\rm max}$ is 
a constant, independent of the galaxy type.
However, as Refs.~\cite{Boyarsky:2009rb,Boyarsky:2009af} have demonstrated this conclusion has been 
based on a too small range of dynamical masses.
When taking into account all type of DM halos, from dSph galaxies to galaxy clusters, one can see 
that $\Sigma_{\rm max}$ (or quantities that can be related to it) increases with $M_{200}$.
As demonstrated in~\cite{Boyarsky:2009af} this scaling can be explained within the secondary infall model.
This result 
has later been confirmed with larger datasets (see e.g.
\cite{Cardone:2010jb,Napolitano:2010cq,DelPopolo:2012eb}) with a scaling
relation given by \cite{DelPopolo:2012eb}
\be
\Sigma_{\rm max} \propto M_{200}^{0.20\pm0.05}\,
\label{sigma_scaling}
\ee
when fitting a single power-law to systems (almost) exclusively composed of DM.
Let us stress that we introduced here the {\it maximal} surface density only for the sake of simplicity.
In practice, one would instead choose a radius which is small but where the enclosed mass and hence the 
surface density is observationally still well constrained. As long as this radius is directly proportional to the core 
radius, or the scale radius in the case of an NFW profile, this does not affect the scaling with $M_{200}$ 
(but can significantly reduce the observational scatter in this relation~\cite{Boyarsky:2009rb,Boyarsky:2009af}).

\begin{figure}[t]
  \centering
  \includegraphics[width=0.9\textwidth]{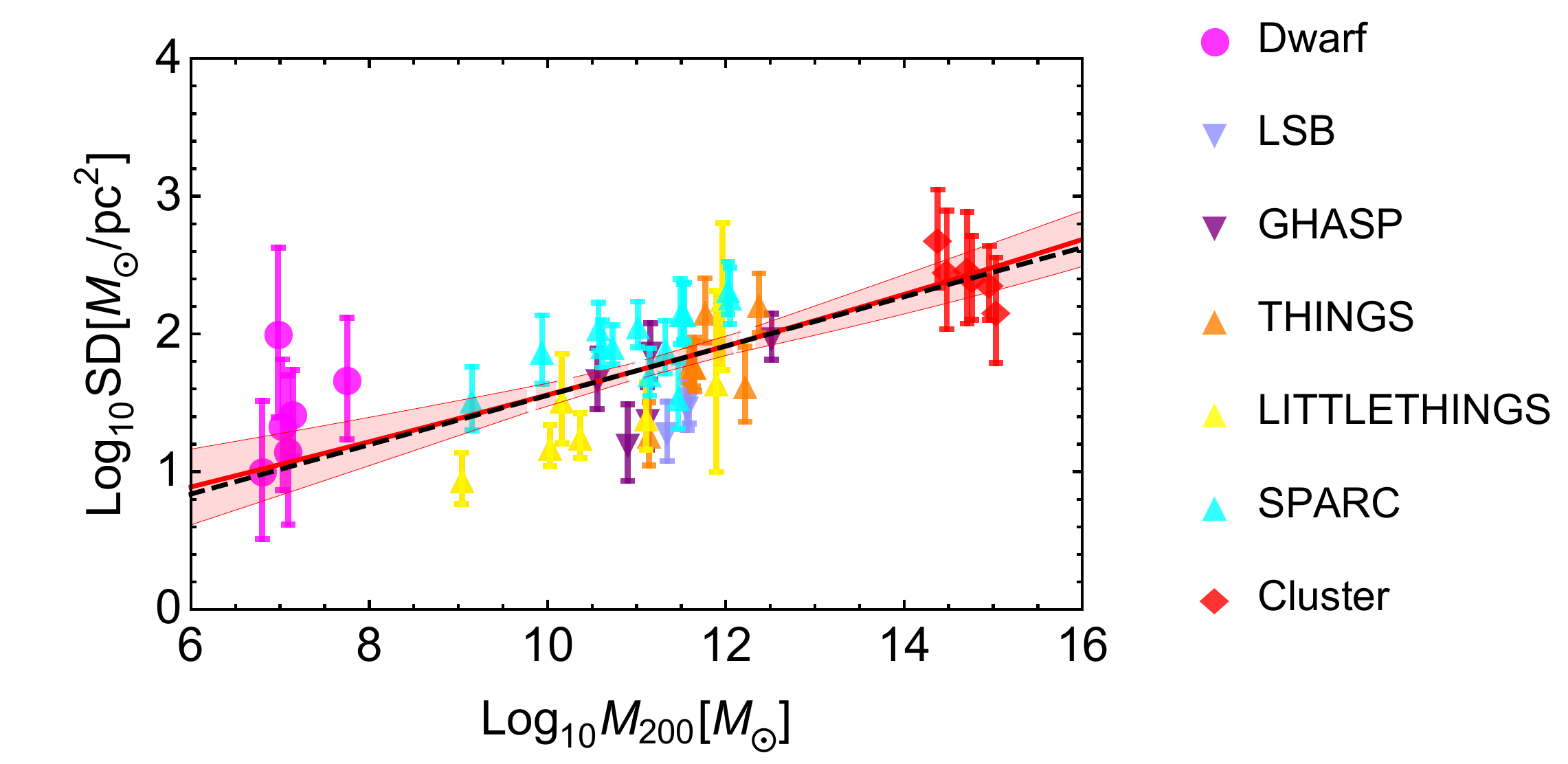}
    \caption{Surface density $\Sigma$ at the scale radius of the NFW profile, as a function of the halo mass $M_{200}$, for the 
    objects described in more detail in Appendix \ref{app:data}. The dashed black line indicates the best fit to 
    a simple scaling relation, $\Sigma\propto M_{200}^n$ as given in 
    Eq.~(\ref{sigma_scaling_NFW}). The red solid line corresponds instead to the best fit of a power-law in the 
    concentration-mass relation as expected in $\Lambda$CDM cosmology, c.f.~Eq.~(\ref{our_cM}), with  the shaded 
    region indicating the result of a 2$\sigma$ variation in normalization and and slope of this relation.}
    \label{fig:SDNFW}
\end{figure}

For the purpose of constraining DM self-interactions as explained further down, we selected DM dominated objects 
over a large range of halo masses where we could find reasonable fits to both  a cored profile and an NFW profile 
in the literature (for more details, see Appendix~\ref{app:data}). 
For an NFW halo, we have 
\be
\label{MNFW}
 M_{\rm NFW}(r)=4\pi\rho_s r_s^3\left[
 \log\left(\frac{r_s+r}{r_s}\right)-\frac{r}{r+r_s}
 \right]\,,
\ee
and hence
\be
\Sigma_{\rm NFW} (r_s)= \frac32\left(\log4-1\right) \rho_s r_s\approx0.579\, \rho_s r_s\,.
\ee
In Fig.~\ref{fig:SDNFW}, we show the surface densities of the objects in our sample, inferred from the fit 
to the NFW profile and evaluated at the scale radius. The errors of the 
data points we calculate by using uncorrelated 1$\sigma$ errors on the NFW parameters quoted in the 
corresponding references (using instead directly the kinematic constraints on the parameter combination 
$\rho_s r_s$, as available for dwarf galaxies, would result in much smaller errors). We then fit a power-law to these 
data points, resulting in 
\be
\Sigma_{\rm NFW} (r_s) = 0.58^{+0.50}_{-0.27}\, \left(\frac{M_{200}}{M_\odot}\right)^{0.179\pm 0.024} 
\text{ M}_{\odot}/\text{pc}^2\,,
\label{sigma_scaling_NFW}
\ee
which is consistent with the scaling reported earlier. We indicate the best-fit power-law as a black dashed
line in Fig.~\ref{fig:SDNFW}.

\subsection{$\Lambda$CDM interpretation}

To proceed, it is useful to introduce the halo {\it concentration},
\be
 c\equiv\frac{r_{200}}{r_s}\,,
\ee
where $r_{200}$ is the virial radius defined as the radius inside which the mean density
$\langle\rho\rangle$ exceeds the critical density $\rho_c=3H^2_0/(8\pi G)$ by a factor of 
200 (for other profiles than NFW, suitable generalizations of this definition of $c$ exist \cite{Ludlow:2013vxa}), i.e.
\be
\label{eq:r200}
r_{200}\equiv \left(\frac{G}{100H_0^2} \right)^\frac13 M_{200}^{1/3}
=1.62\cdot10^2\,\left(\frac{M_{200}}{10^{12} h^{2} M_\odot}  \right)^{\frac13}{\rm kpc}\,.
\ee
This implies 
\be
\rho_s\equiv\rho_c \,\delta= \rho_c \frac{200}{3}\frac{c^3}{\log(1+c)-c/(1+c)}\,,
\ee
which allows us to exchange the parameters $(\rho_s,r_s)$ for $(M_{200},c)$. For the surface density
inside the scale radius, we thus find 
\bea
\Sigma_{\rm NFW} (r_s)&=&\frac{3(\log4-1)}{8\pi}\left(\frac{100H_0^2}{G}\right)^\frac23\,
M_{200}^\frac13\,
\frac{c^2}{\log(1+c)-c/(1+c)}\\
&=&1.74\,h^2\left(\frac{M_{200}}{10^{12} h^{2} M_\odot}  \right)^{\frac13}\frac{c^2}{\log(1+c)-c/(1+c)}
\, \frac{M_\odot}{{\rm pc}^2}\,.
    \label{eq:sigma_cm}
\eea

From numerical simulations \cite{Maccio:2006wpz,Neto:2007vq,Maccio:2008pcd,MunozCuartas:2010ig,2012MNRAS.423.3018P,Klypin:2014kpa,Pilipenko:2017iae}, 
but also observations \cite{Comerford:2007xb,Sereno:2013aod,Merten:2014wna,Du:2015dua,Groener:2015cxa,Amodeo:2016wtq,Cibirka:2016nhw,Umetsu:2015baa,Lieu:2017xkq}, the concentration is not 
independent of the halo mass, but rather follows a simple scaling law. There is, however, a significant object-to-object 
scatter associated to this relation, and even the best-fit values for slope and normalization of this
power-law differ in the literature. One of the
more often used results is the one by Macci\`o {\it et al.} \cite{Maccio:2008pcd} 
\be
    c = 8.3
    \left(\frac{M_{200}}{10^{12}h^{-1}M_{\odot}}\right)^{-0.104}.
    \label{eq:cmass}
\ee
Using the mean value of this slope in Eq.~(\ref{eq:sigma_cm}) results in the slope $d(\log \Sigma_{\max})/d(\log M_{200})$
ranging from roughly 0.13 at low masses to 0.25 at very large masses. In $\Lambda$CDM cosmology, the scaling 
of the surface density as given in  Eq.~(\ref{sigma_scaling}) can thus consistently be interpreted as a reflection of 
the above concentration-mass relation \cite{Boyarsky:2009rb,Cardone:2010jb,Napolitano:2010cq,DelPopolo:2012eb}. 

We note that, from Eqs.~(\ref{eq:sigma_cm},\ref{eq:cmass}), we can infer not only the scaling of 
$\Sigma$ with $M_{200}$, but also the normalization of that relation in $\Lambda$CDM 
cosmology. In fact, we can turn the argument around, and provide an updated {\it measurement} of the 
concentration-mass relation by fitting Eq.~(\ref{eq:sigma_cm}) to the data points shown in 
Fig.~\ref{fig:SDNFW}. Assuming again a simple power-law for $c(M_{200})$, we find 
\be
c = (10.8\pm 0.6)\, \left(\frac{M_{200}}{10^{12}h^{-1}M_{\odot}}\right)^{-0.103\pm 0.015}\,.
\label{our_cM}
\ee
In Fig.~\ref{fig:SDNFW}, we show the resulting surface density as solid red
line, with the shaded red region indicating the uncertainty in the concentration-mass relation 
that we derived.

\subsection{Cored halo profiles}
\label{sec:cored_profiles}
Let us now discuss how the scaling of the mean surface density would change in the presence
of cored profiles as produced by SIDM.
 For small core radii, much smaller than 
the scale radius $r_s$ of the NFW profile expected in the outer parts, the enclosed mass 
inside $r_s$ will obviously not be significantly affected. We are hence no longer interested in 
evaluating the surface density at $r_s$, as before, but at a smaller radius. The optimal choice
in this respect is the core radius itself.\footnote{%
We note that the surface density at the core radius was previously considered in the context 
of cores produced by strong DM self-{\it annihilation}, resulting in a scaling relation formally 
independent of the annihilation rate in this limit~\cite{Lin:2015fza}. 
It was then suggested that the same could be expected for DM 
{\it self-interactions}~\cite{Lin:2015fza,Kamada:2016euw}. Our results show that this is {\it not}
the case: a surface-density independent of the self-interaction cross-section 
is not supported by numerical $N$-body simulations.
}
While the surface density of NFW and cored 
profiles would differ even more at radii smaller then the core radius, in particular, the surface 
density in this regime is much less well constrained by observational data (and would anyway 
{\it decrease} with respect to its value at the core radius). 
We use this opportunity to stress again that the scale of efficient self-interactions, $r_{\rm SIDM}$, 
is essentially impossible to determine observationally, so we are also not interested in 
considering the surface density at this (somewhat larger) radius.

For any cored profile parameterization, we could now in principle follow an approach in analogy to
what we did for the NFW case, and analytically express $\Sigma(r_0)$ in terms of the halo mass
and the core scale parameter $r_0$. This, however, is impractical because it would have to be done for each 
choice of profile parameterization separately. Besides, one would need independent prescriptions of how to
relate $r_0$ to the scale of efficient self-interactions, $r_{\rm SIDM}$, for each of these cases, which 
complicates the analysis we are interested in here.
In the spirit of keeping the discussion as model-independent and general as possible,
we therefore consider instead the {\it experimental core radius} $r_c$, which we define as the radius 
at which the best-fit NFW profile equals the central density of the best-fit cored profile,
\begin{equation}
    \rho_{\text{NFW}} (r_c) \equiv \rho_{\text{cored}}(0)\,.
    \label{eq:rcdef}
\end{equation}
Unlike the profile parameter $r_0$ that appears, e.g., in the Burkert profile and the modified isothermal sphere, 
the core radius $r_c$ defined in this way is relatively independent of which cored profile is used for the fit, 
and typically more robustly constrained observationally.  
The ``observed'' surface density at this core radius is then simply given by
\be
 \Sigma_{c,\,\mathrm{obs}} = \langle \rho_{\text{cored}} \rangle_c\, r_c\,,
 \label{SDrcdef}
\ee 
where we note that $ \langle \rho_{\text{cored}} \rangle_c \approx \rho_{\text{cored}}(0)= \rho_{\text{NFW}} (r_c)$
because the DM density inside $r_c$ is almost constant (we do not use this last approximation in our
analysis).

\begin{figure}[t]
  \centering
   \includegraphics[width=0.38\textwidth]{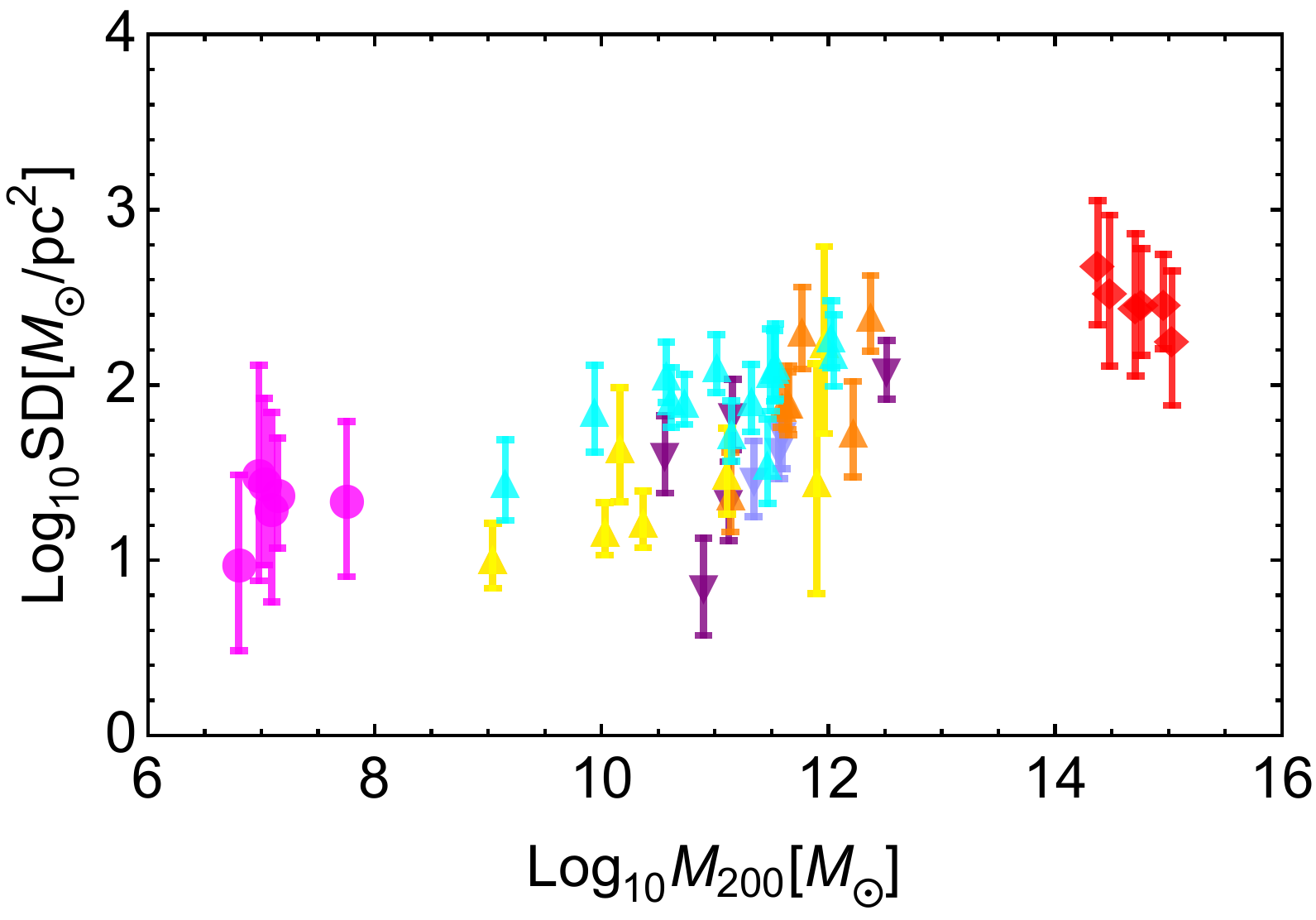}
   ~\includegraphics[width=0.57\textwidth]{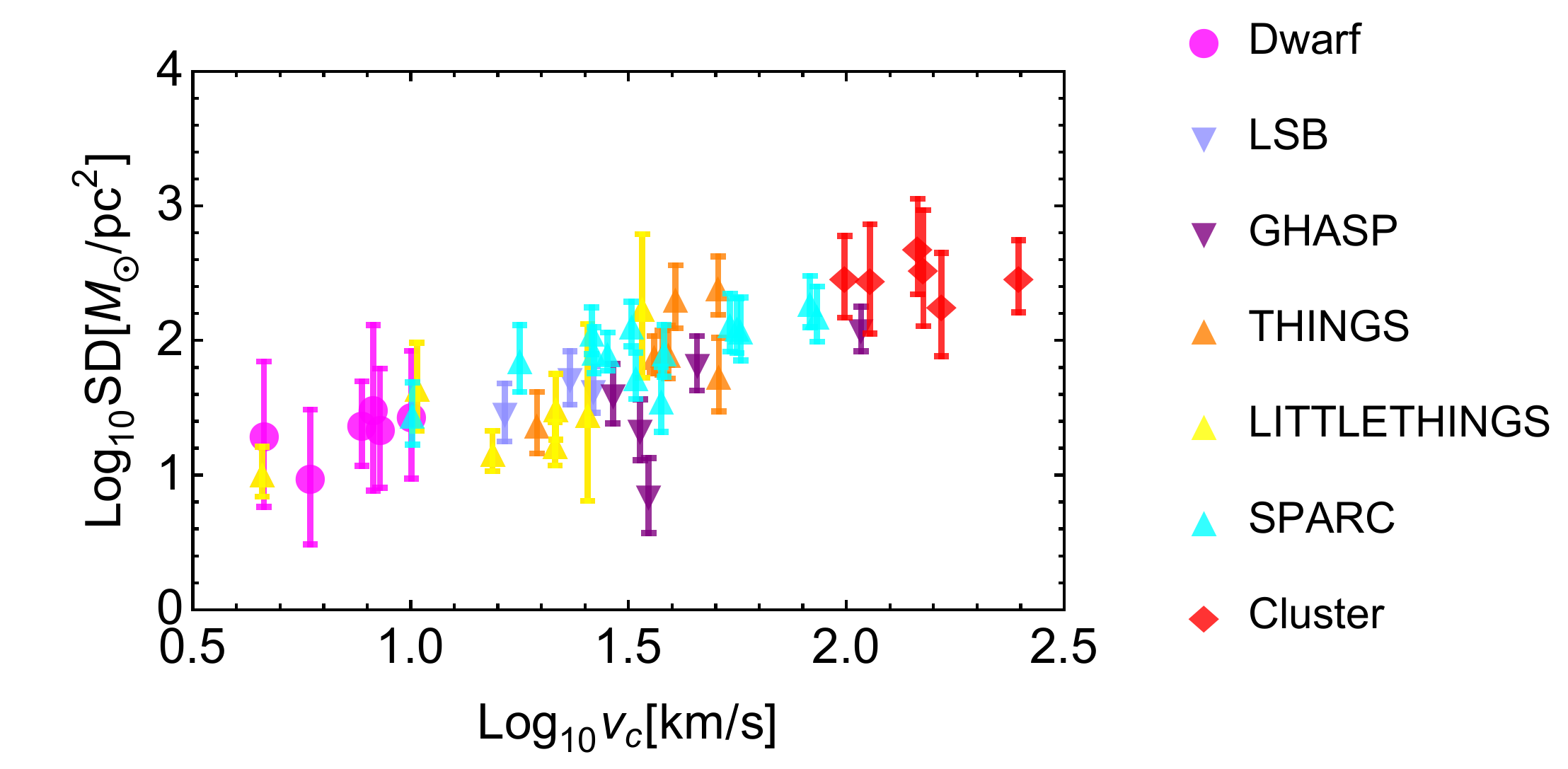}
    \caption{The dependence of the surface density at the experimental core radius $r_c$, as function of $M_{200}$ and the circular velocity $v_c$ at this radius, for the 
    objects described in more detail in Appendix~\ref{app:data}.}
    \label{fig:SDexpdata}
\end{figure}

We plot  $\Sigma_{c,\,\mathrm{obs}}$ in Fig.~\ref{fig:SDexpdata}, for the same objects that 
we used in Fig.~\ref{fig:SDNFW}, as a function of both virial mass and the circular velocity at the core radius.
Concretely, we assumed that the parameters of the NFW and cored profile 
follow a Gaussian distribution in each case, with standard deviation as quoted in Appendix \ref{app:data}.
Drawing parameters from this distribution, we then determined $\langle \rho\rangle_c$ and 
$r_c$ for a large sample of  profile realizations  for each halo, to calculate $v_c(r_c)$ and 
$\Sigma_{c,\,\mathrm{obs}}$. The data points in Fig.~\ref{fig:SDexpdata} 
thus created provide the surface density corresponding to the best-fit profile parameters, with
errors corresponding to one standard deviation in our sample.
We note that this Monte Carlo approach results in conservative estimates for the errors 
$\Delta \Sigma_{c,\,\mathrm{obs}}$, because it treats the two profile 
fits to identical halos as being independent. This will result in conservative limits when eventually 
using this dataset to constrain the DM self-interaction rate.

\begin{figure}[t]
    \centering
    \includegraphics[width=0.48\linewidth]{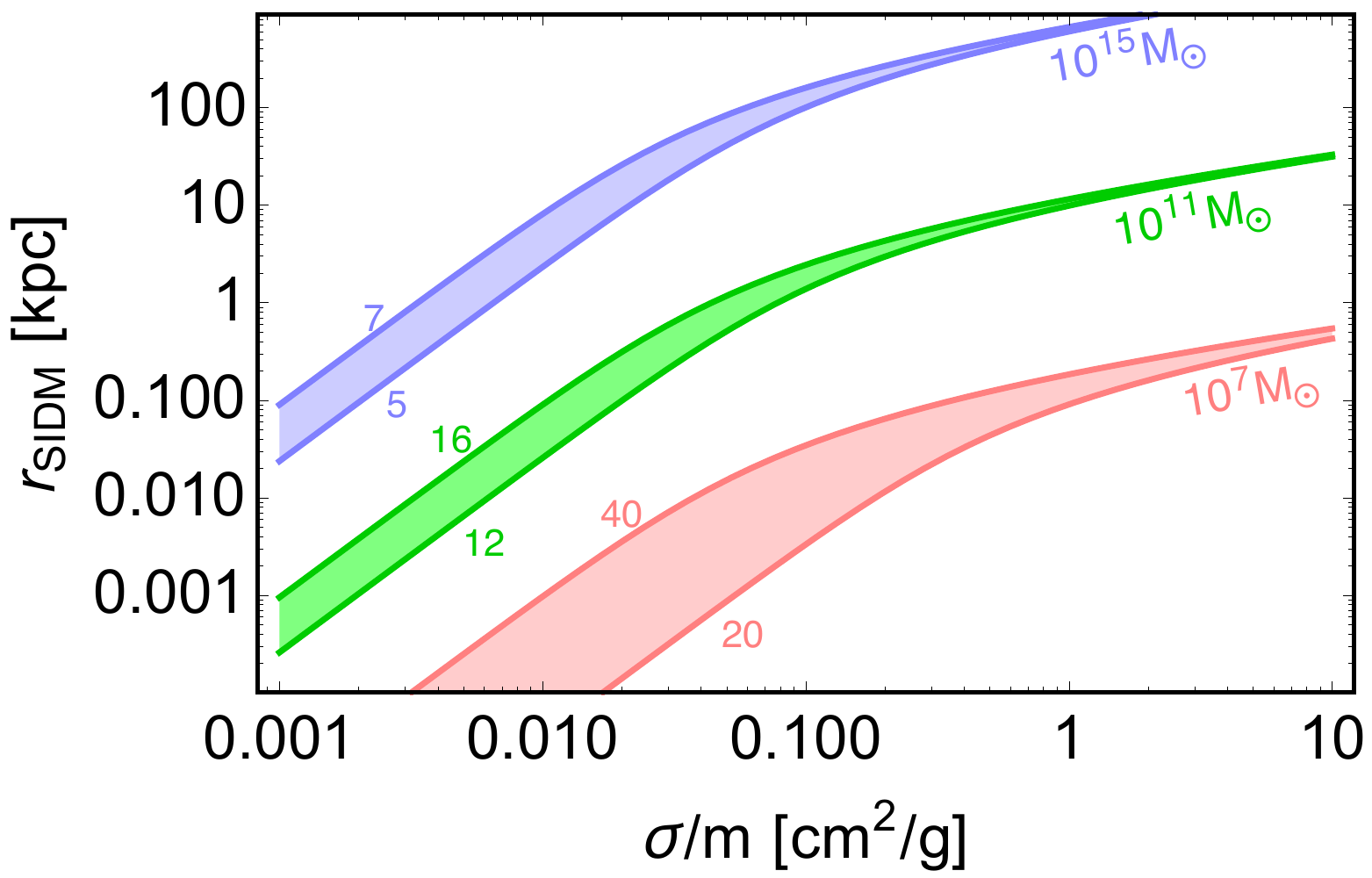}
    ~~\includegraphics[width=0.48\linewidth]{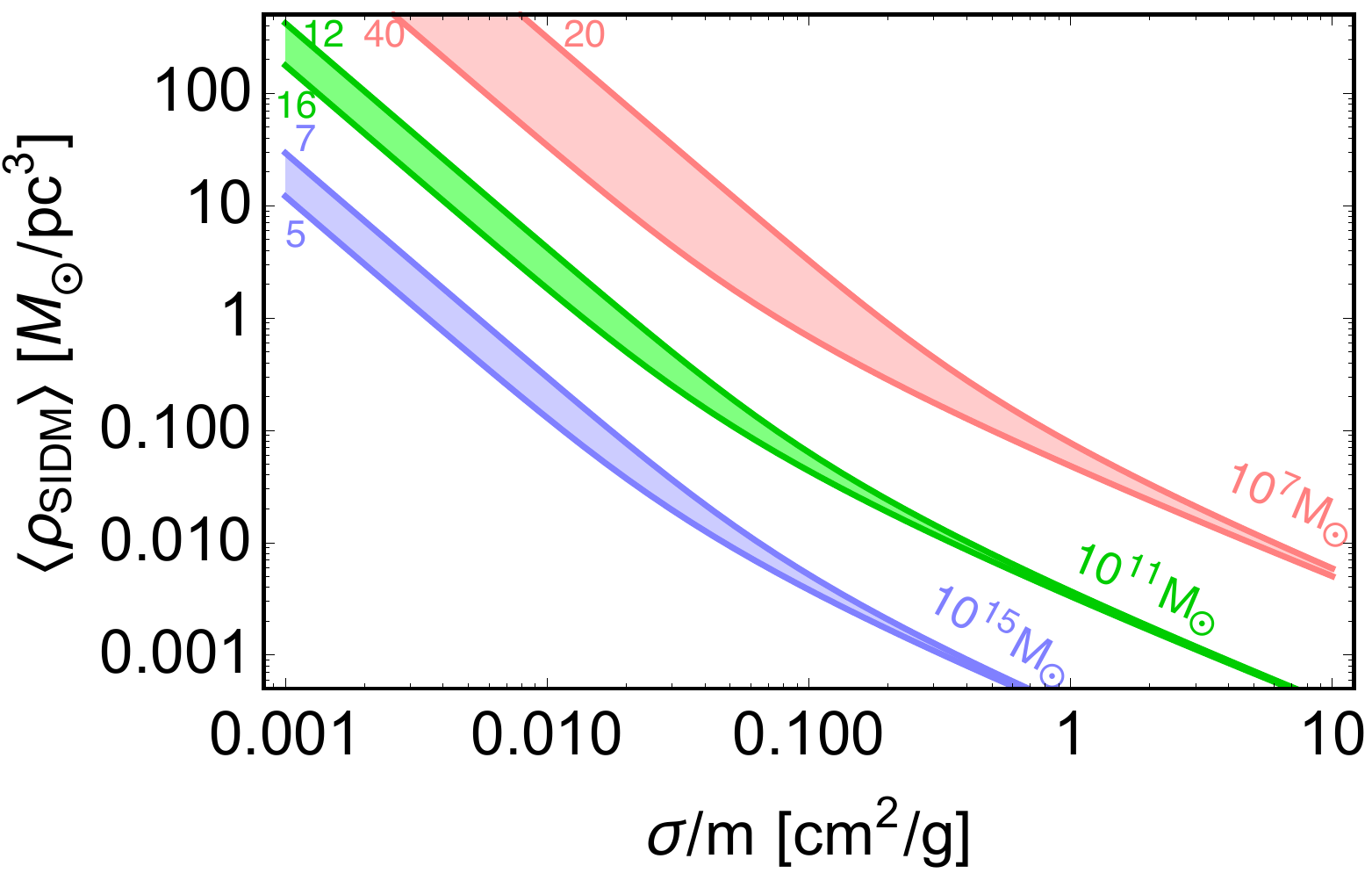}
    \caption{
    $r_{\text{SIDM}}$ and $\langle\rho\rangle_{\text{SIDM}}$ as a function of the self-interaction 
    cross section for three different 
    halo masses $M_{200}=10^7\,M_{\odot}$ (red), $10^{11}\,M_{\odot}$ (green) and $10^{15}\,M_{\odot}$ (blue). For 
    each halo mass we show the result of varying the halo concentration,  within typical values of $c$ as indicated 
    next to the envelopes of those shaded regions. For the purpose of this figure, we take 
    $\alpha_{\rm SIDM}=2$ and $\xi = 1$.}
    \label{fig:rrhoSIDM}
\end{figure}

Let us now consider the theoretically expected surface density at the core radius, assuming that the
existence of the core is the result of DM self-interactions. We start by recalling that for $r>r_{\rm SIDM}$ we expect the 
standard NFW profile to be unaffected by DM self-interactions. This implies that the mass, and hence 
the average (surface) densities inside $r_{\rm SIDM}$ remain the same:
\begin{equation}
    \langle\rho\rangle_{\text{SIDM}} = 
    \langle\rho_{\text{NFW}}\rangle_{r_\text{SIDM}}=
    \frac{M_\mathrm{NFW}(r_\text{SIDM})}
    {({4\pi}/{3})\, r_\text{SIDM}^3}\,.
    \label{eq:avrho}
\end{equation}

We can then use $M_\mathrm{NFW}$ from Eq.~(\ref{MNFW}) and the implicit definition of $r_{\rm SIDM}$ given 
in Eq.~(\ref{rsidmimp}) to relate the self-interaction 
scale to the scale radius, for a given NFW profile and self-interaction cross section:
\be
  r_{\text{SIDM}} = \eta\, r_s\,,
  \label{etadef}
\ee
where $\eta$ is a solution to the equation
\bea
    \left( \ln(1+\eta) -\frac{\eta}{1+\eta}\right) \eta^{-7/3}&=&
    \left(
    {\frac{576}{\alpha_{\text{SIDM}}} G \left(\dfrac{\sigma}{\xi m_\chi}\right)^2 t_{\text{age}}^2 \rho_s^3 r_s^2}\right)^{-1/3}.
   \label{eq:etaresult}
\eea
For illustration, we show in Fig.~\ref{fig:rrhoSIDM} how the resulting $r_{\text{SIDM}}$ the average density $\langle\rho\rangle_{\text{SIDM}}$  
(from Eq.~(\ref{rsidmimp})) scale with the self-interaction strength. In order to produce these curves, we take 
$\xi=1$, $\alpha_{\text{SIDM}}=2$ and implement an average halo ago $t_{\rm age}$ following the
prescription in Appendix~\ref{app:age}. The range of concentrations for each mass displayed in this figure
very roughly corresponds to that given by the $c-M_{200}$ relation, see Eq.~\eqref{our_cM}. For the displayed 
halo mass of $10^7~M_{\odot}$, we allow for a larger scatter taking into account that dwarf 
satellites are not too well fitted by this relation, see Fig.~\ref{fig:SDNFW} and the 
discussion in Refs.~\cite{Boyarsky:2009rb,Boyarsky:2009af}.

Once we have $\langle\rho\rangle_{\text{SIDM}}$, we obtain the theoretically expected surface density as
\be
 \Sigma_{c,\,\mathrm{theo}} = \langle \rho\rangle_c r_c=
 \kappa r_c \langle\rho\rangle_{\text{SIDM}}\,.
  \label{eq:SDthnew}
\ee
In the last step we have introduced a phenomenological ratio $\kappa$
of the average densities inside $r_\mathrm{SIDM}$ and $r_c$, respectively:
\begin{equation}
    \kappa  \equiv \frac{\langle\rho\rangle_{c}}{\langle\rho\rangle_{\text{SIDM}}}\,.
    \label{eq:kappa}
\end{equation}
This ratio, as argued in Section \ref{sec:cores}, is difficult to
determine observationally or directly from first principles. 
On the other hand it is a quantity that turns out to be tightly constrained by 
simulations, and this is what we will make use of when determining 
limits on $\sigma/m_\chi$ in Section \ref{sec:constraints}.
In the following, we will simply treat $\kappa$ as a constant, i.e.~independent of cross section
and halo mass, which is consistent with the simulation data we have explicitly 
looked at. We note that the exact form of $\kappa$, and hence the robustness of our 
limits, can be improved by taking into account a larger sample of (new) simulations, which 
is beyond the scope of the present work.

 Let us stress that $\langle\rho\rangle_c$ here
refers to the average DM density inside the same observationally determined core radius $r_c$ 
as defined in Eq.~(\ref{eq:rcdef}). If baryonic effects also contribute to the observed core, then
the net effect of an increasing core size $r_c$ and a decreasing core density $\langle\rho\rangle_c$
is a surface density $\Sigma_{c,\,\mathrm{obs}}$ that is necessarily smaller than the
theory expectation for a halo consisting exclusively of SIDM as stated in Eq.~(\ref{eq:SDthnew}).
This will allow us to place {\it upper}, but not lower limits on the self-interaction rate.
We also note that the above definition of $\kappa$ -- 
via $\langle\rho\rangle_c\approx \rho(0)= \rho_{\text{NFW}} (r_c)$  -- 
implicitly fixes the ratio of $r_c$ and $r_{\text{SIDM}}$ (as a function of $\sigma/m_\chi$, $\kappa$ 
and the NFW profile parameters).
Using $\kappa$ from Eq.~\eqref{eq:kappa}, we can thus also calculate the logarithmic slope $\alpha_{\text{SIDM}}$
of the density profile at $r_{\text{SIDM}}$ that appears in Eqs.~\eqref{rsidmimp} and \eqref{eq:etaresult}. 
To do so, we numerically solve  the Jeans equation~\eqref{eq:Jeanssimp} with $\rho'(0)=0$ 
and determine $\rho_\chi'$ at $r=r_{\text{SIDM}}$, as a function of $\kappa$. 
As explained in Section \ref{sec:cores}, the cored solution to the Jeans equation with fixed central density $\rho_{\chi}(0)$ depends only on one 
dimensionless parameter, 
\mbox{$r_{\text{SIDM}}\sqrt{G\rho_{\chi}(0)}/\sigma_v\equiv r_{\text{SIDM}}/r_{\text{Jeans}}$},
so this mapping between $\alpha_{\text{SIDM}}$ and $\kappa$ must be unique. 

Our formula~\eqref{eq:SDthnew} describes, as physically expected in the weakly interacting regime, 
a surface density that decreases with increasing cross section. This implies that it must break down
once we leave this regime, and the core size instead should start to {\it decrease} again as the profile 
approaches the isothermal $r^{-2}$ solution. We will not consider such large cross sections in our analysis.
In the opposite limit of $\sigma/m_\chi \to 0$, on the other hand, we recover the NFW expectation for 
the surface density at any given radius $r_c$. This is an important property of our model when 
constraining the self-interaction strength $\sigma/m_\chi$ in the next section.


\section{Constraining dark matter self-interactions}
\label{sec:constraints}

\subsection{Statistical  treatment}
\label{sec:stat}

In the previous section, we have discussed how DM self-interactions can affect the observed 
scaling of the surface density with halo mass. Here, we will derive constraints on the
DM self-scattering cross section based on this prescription. As experimental input for our 
analysis, we consider the surface density at the experimental core radius $r_c$ for
the objects described in Appendix~\ref{app:data}, as shown in Fig.~\ref{fig:SDexpdata}. 
In order to determine limits on the ``signal'' strength, 
\be
S = \frac{\tilde\sigma}{m_\chi}\equiv \frac{\sigma}{\xi m_\chi}\,,
\label{sigeff}
\ee
we then use the likelihood ratio test \cite{Rolke:2004mj}. Here, we have introduced an 
{\it effective cross section} $\tilde\sigma$ to reflect the degeneracy of the physical 
scattering cross section $\sigma$ with the $\mathcal{O}(1)$ factor $\xi$ as introduced in our
basic SIDM ansatz of Eq.~(\ref{sidm_ansatz}).
We model the total likelihood as a product of normal distributions over each halo object (``data point'') $i$,
\be
\mathcal{L}=\Pi_i\, N(f_i|\mu_i,\sigma_i)\,,
\ee
where $f_i$ is the value of the (logarithmic) surface density inferred from the kinematical anlaysis,
$\sigma_i$ its variance, and $\mu_i$ is the (logarithmic) surface density predicted by the model.
The latter depends not only on the self-interaction strength $S$, but in principle also on a number of 
nuisance parameters $\{\alpha_k\}$.
The contribution from DM self-interactions therefore enters with a single, non-negative 
degree of freedom, which implies that 95\%\,CL upper limits on $S$ are derived by increasing 
$S$ from its best-fit value until $-2 \ln \mathcal{L}$ has changed by 2.71, while re-fitting (``profiling over'') 
all nuisance parameters $\{\alpha_k\}$. 

Concretely, we use Eq.~(\ref{SDrcdef}) for the data points $f_i$ and Eq.~(\ref{eq:SDthnew}) for the model
prediction $\mu_i$. For the variance, we add observational and theory uncertainties in quadrature,
$\sigma_i^2= \sigma_{i,{\rm obs}}^2+ \sigma_{i,{\rm theo}}^2$. The errors in the ``observed'' surface density 
uncertainty, $\sigma_{i,{\rm obs}}=\Delta \Sigma_{c,\,\mathrm{obs}}$, are determined as described in 
the previous Section and indicated in Fig.~\ref{fig:SDexpdata}. In the  ``theory error'' in this figure, 
we include two contributions:
\be
\sigma_{i,{\rm theo}}^2= \sigma_{i,{\rm halo}}^2  + \sigma_{t_{\rm age}}^2\,. 
\ee
Here, the largest contribution, $\sigma_{i,{\rm halo}}$, is related to uncertainties in the observational 
determination of the halo parameters and picks up two contributions: {\it i)} from the variance in the 
core radius $r_c$, which is determined in the same way as for $\sigma_{i,{\rm obs}}$ and directly 
enters in the model prediction $\mu_i$ via Eq.~(\ref{eq:SDthnew}), and {\it ii)} from error propagation 
of the NFW halo parameters ($\rho_s$ and $r_s$) that enter in Eq.~(\ref{eq:etaresult}) for the 
average density $\langle\rho\rangle_{\rm SIDM}$. For the halo age, we adopt a value of
$\sigma_{t_{\rm age}}$ that corresponds to a rather generous factor of 1.2~\cite{carlos_private} in the expected
halo-to-halo scatter of the implemented $t_{\rm age}(M_{200})$ relation described in Appendix \ref{app:age}.
The value of $\kappa$ in Eq.~(\ref{eq:SDthnew}), finally, we extract from direct comparison 
with SIDM simulations~\cite{Rocha:2012jg},\footnote{
Concretely, we consider 6 halos from these SIDM simulations with a self-interaction cross section 
$\sigma/m=1\text{ cm}^2/\text{g}$ and identical CDM initial conditions, with halo masses between 
$10^{10}$ and $2\cdot10^{14}~M_{\odot}$. Using the simulation data, we determine $r_c$ in the
standard way, as in Eq.~(\ref{eq:rcdef}), and $r_{\text{SIDM}}$ as the radius when the velocity dispersion 
starts to differ from the  CDM expectation by 10\%.}
finding $\langle\kappa\rangle = 3.9$ with a scatter of 
$\sigma_{\kappa}^2\equiv  \langle\kappa^2\rangle -\langle\kappa\rangle^2 = 1.4^2$. 
We vary $\kappa$ freely within this range.

\subsection{Results}
\label{sec:res}

\begin{figure}[t!]
  \centering
    \includegraphics[width=0.32\textwidth]{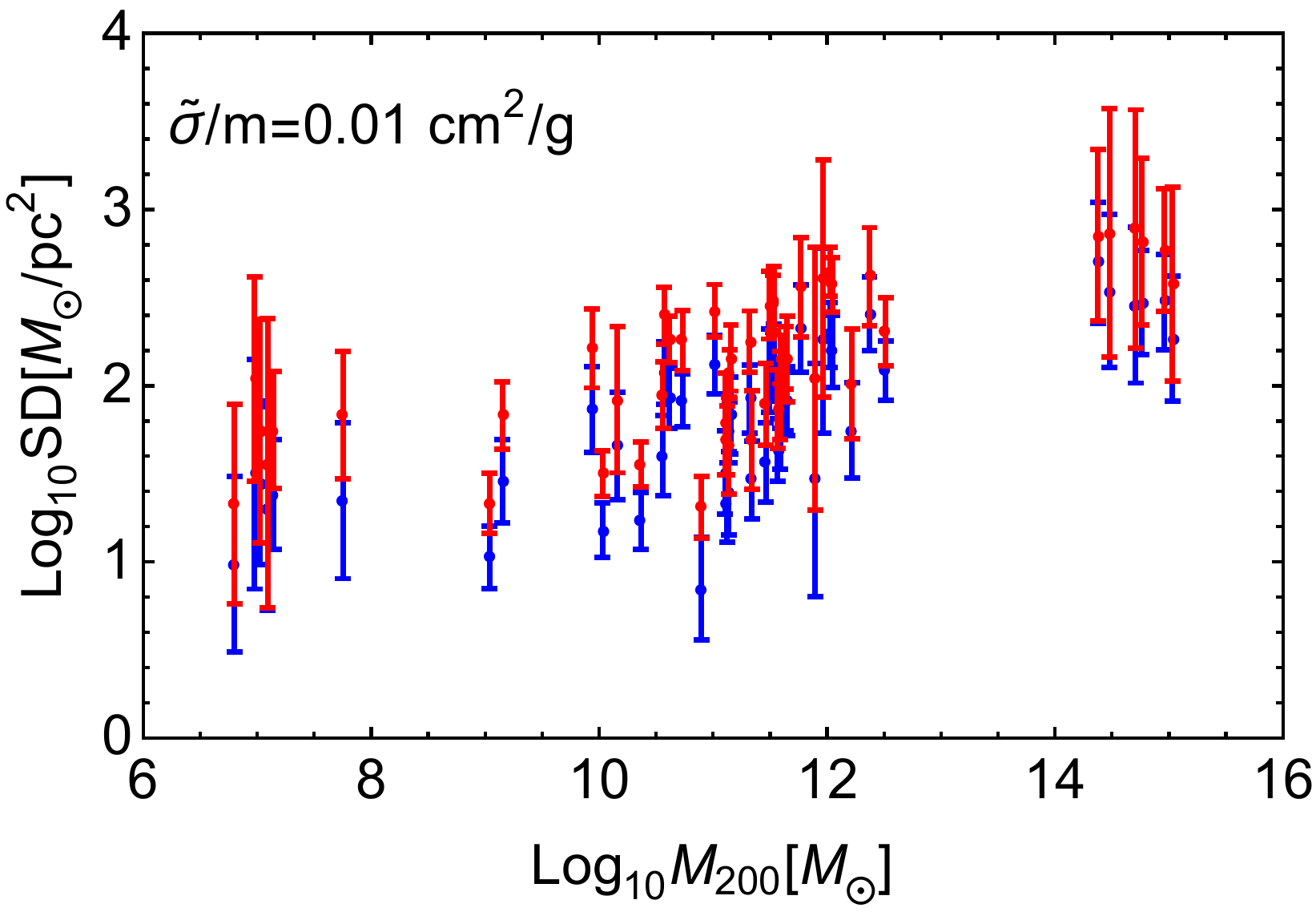}
  \hfill
    \includegraphics[width=0.32\textwidth]{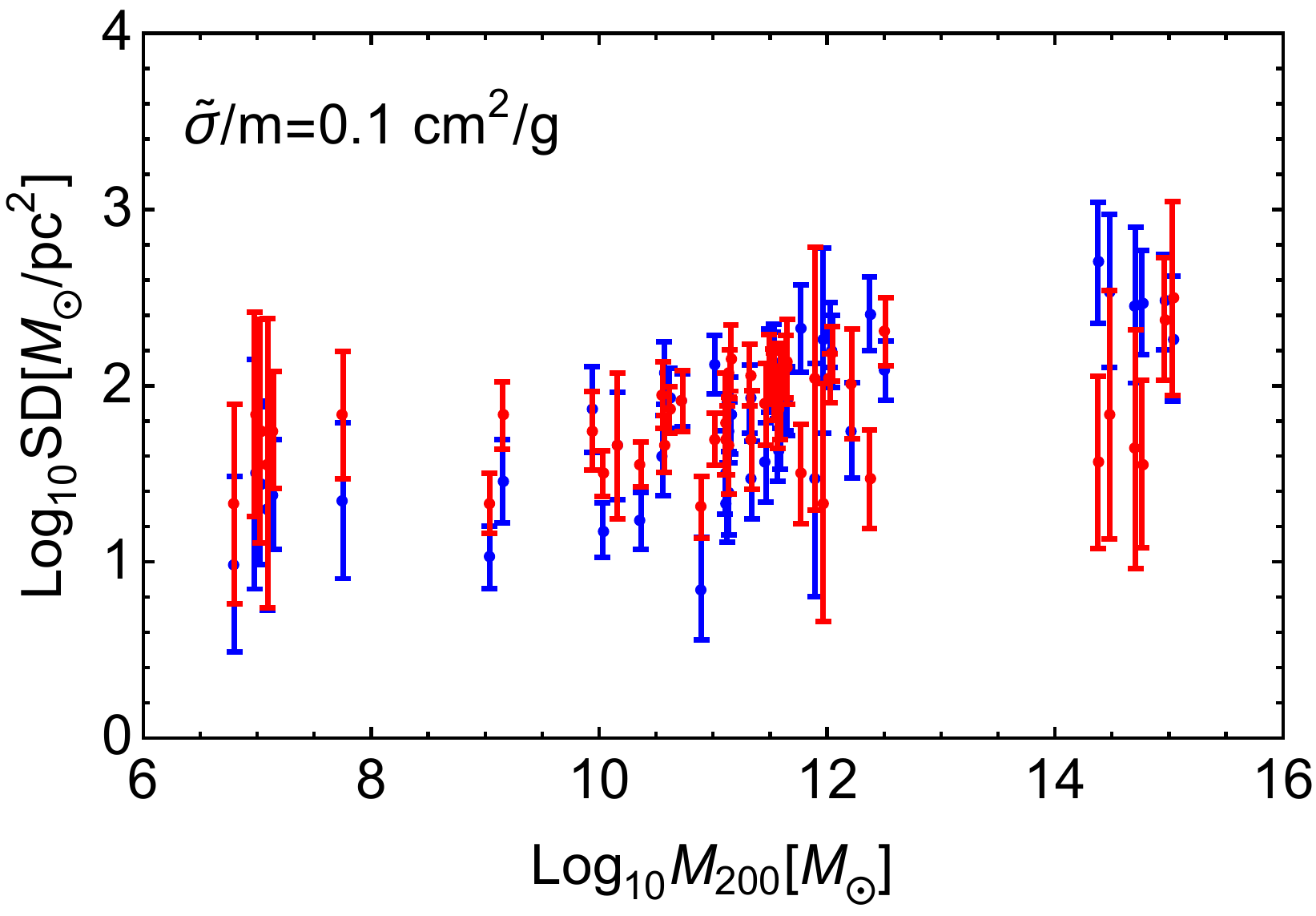}
\hfill
   \includegraphics[width=0.32\textwidth]{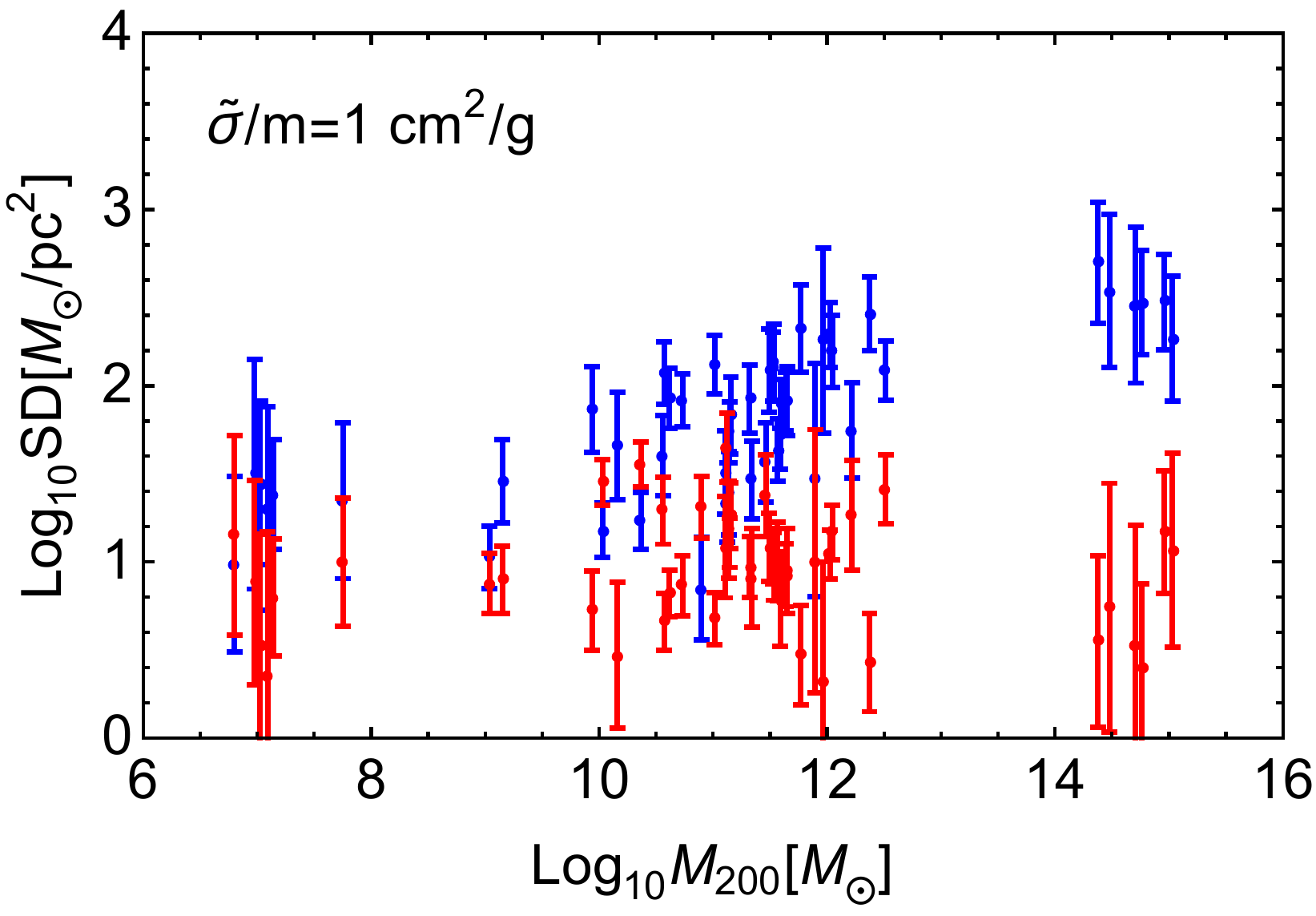}
    \caption{The dependence of the surface density at the experimental core radius $r_c$, as function of the 
     halo mass, for the objects described in more detail in Appendix \ref{app:data}. The blue points represent the 
     observational input, as in Fig.~\ref{fig:SDexpdata}.
    The red points show the theory expectation for SIDM, Eq.~(\ref{eq:SDthnew}), with errors as described in 
    detail in Section \ref{sec:stat} (for the sake of illustration, we here use a fixed value of $\kappa=3.9$). 
     From left to right, 
    the panels show the results for an effective scattering
    cross section, c.f.~Eq.~(\ref{sigeff}), of $\tilde \sigma/m_{\chi}=0.01$\,cm$^2$/g, $0.1$\,cm$^2$/g and
    $1$\,cm$^2$/g, respectively.}
    \label{fig:theoryexp2}
\end{figure}

In order to illustrate our approach, we compare in Fig.~\ref{fig:theoryexp2} the observed surface density
(as shown in Fig.~\ref{fig:SDexpdata}) with the SIDM predictions that we derived above, for various
values of the effective self-interaction cross section. These plots clearly 
demonstrates that a too large self-interaction cross section would be inconsistent with the 
experimental data. In fact, we even see a slight preference for a non-zero value of $\tilde\sigma/m_\chi$.
Of course, the latter cannot be taken as an indication for SIDM, but is rather a reflection of the fact that we
consider objects consistent with a cored profile -- which leads to a smaller surface density than what 
would be expected for the NFW case.

\begin{figure}[t!]
  \centering
    \includegraphics[width=0.9\textwidth]{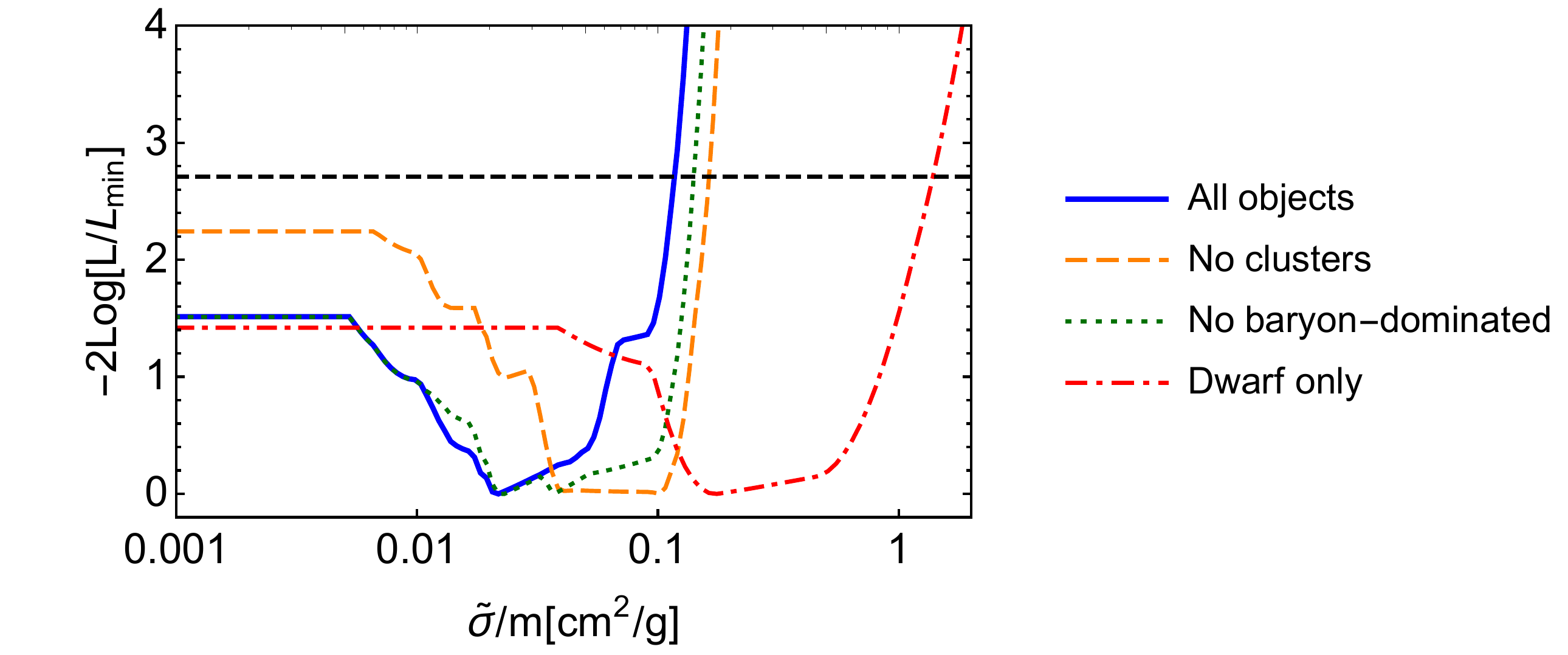}
    \caption{The log-likelihood ratio as function of the effective cross section $\tilde{\sigma}/m_\chi$ for 
    our full data set (blue solid line) and various subsets: excluding clusters (orange dashed line), 
    excluding objects that are baryon-dominated in the central part (green dotted line), as well as
    the likelihood based {\it only} on dwarf galaxies (red dot-dashed line). The  
    dashed black line indicates the value of the log-likelihood ratio that correspond to a 95\%\,CL upper
    limit. See Section \ref {sec:disc} for a detailed discussion of the various lines.}
    \label{fig:like}
\end{figure}

We make these observations more quantitative by showing in Fig.~\ref{fig:like} the 
full likelihood described in Section \ref{sec:stat} as a function of  $S={\tilde\sigma}/{m_\chi}$.  
From this, we can read off an {\it upper bound} of
\be
  \tilde \sigma/m_\chi \lesssim 0.12\,{\rm cm}^2/{\rm g}, 
  \label{eq:stildemax}
\ee
which roughly corresponds to a 95\%\,{\rm C.L.} limit as we have allowed $\kappa$ to vary up
to its maximal range within $2\sigma$ (allowing values up to $\kappa\leq 8.1$, 
i.e.~within the $3\sigma$ range of $\kappa$, the limit would relax to 
$\tilde \sigma/m_\chi \lesssim 0.14\,{\rm cm}^2/{\rm g}$).
We emphasize again that our method does not allow to put a {\it lower} bound on the cross
section because there are also baryonic feedback processes, not modelled here, which
could lead to a core and hence a reduced surface density.
In order to understand how the above constraint depends on the type of objects that we 
include in our analysis, we include for comparison also the likelihood resulting from various 
subsets of our full halo sample. We will discuss this in more detail in Section \ref{sec:disc} below.

\begin{figure}[t!]
  \centering
   \includegraphics[width=0.7\textwidth]{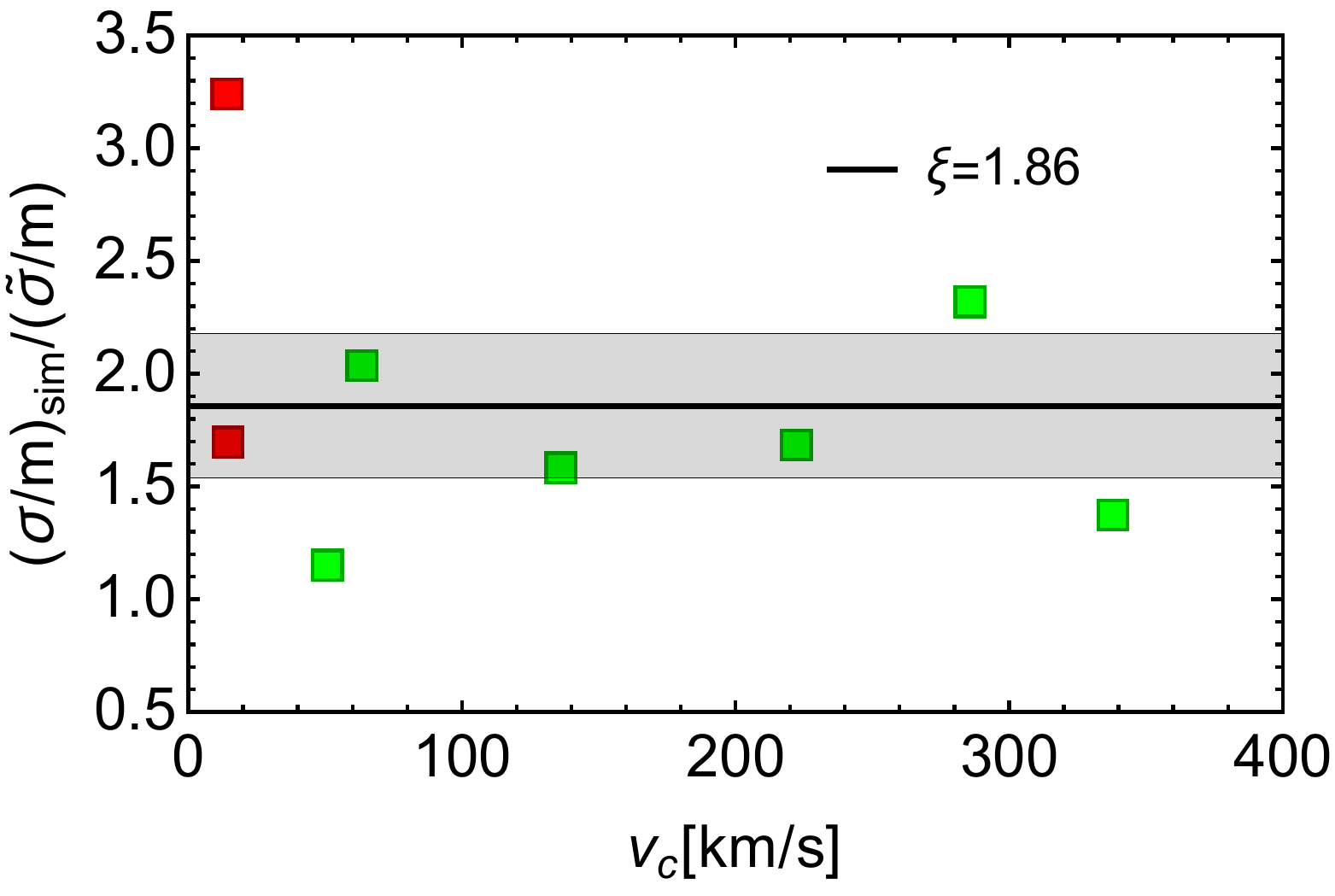}
    \caption{The values of $\xi$ obtained by applying our method to the simulated 
    halos from Refs.~\cite{Rocha:2012jg} (green points) and~\cite{Elbert:2014bma} (red points). 
    The black line is the best fit to the points.}
    \label{fig:rochasigma}
\end{figure}

Lastly, we would like to estimate the systematic bias of our model ansatz, i.e.~the difference between
the effective cross section $\tilde\sigma$ and the physical cross section $\sigma$ that we introduced
in terms of the parameter $\xi$. To do so, we apply the same analysis as above, but now to 8 
{\it simulated} halos with cross section $1\text{ cm}^2/\text{g}$ in the mass range from 
$5\cdot10^{9}~M_{\odot}$ to $2\cdot10^{14}~M_{\odot}$ from Refs.~\cite{Elbert:2014bma,Rocha:2012jg}. 
In Fig.~\ref{fig:rochasigma}, we show the ratio of the reconstructed value of the cross section to the 
``true'' cross section (i.e.~the one used in the simulations). 
For this ratio, we find a best-fit value of 
\be
\xi= \frac{(\sigma/m_\chi)_{\rm sim}}{\tilde\sigma/m_\chi}=1.86\pm 0.32\,.
\label{eq:xi_result}
\ee
Ideally, we would of course use more simulated halos, for cross sections closer to the value 
of roughly $0.1\text{ cm}^2/\text{g}$ that 
corresponds to our constraint, but such simulation results are currently not publicly available. The real theoretical 
uncertainty encoded in the factor $\xi$ (as well as $\kappa$) may thus be 
somewhat larger, but a full investigation of this effect is beyond the scope of this work.
Taking this caveat into account, we arrive at a limit of approximately
\be
  \sigma/m_\chi \lesssim 0.3\,{\rm cm}^2/{\rm g}
\label{eq:smax}
\ee
for the {\it physical} self-interaction cross section.

\subsection{Discussion}
\label{sec:disc}

In the previous section we have derived an upper bound on the {\it effective} self-scattering 
cross section, and stated the result in Eq.~(\ref{eq:stildemax}). 
Let us stress that this bound is {\it not} exclusively driven by the kinematic data, which allowed
us to construct the surface densities shown in Fig.~\ref{fig:SDexpdata}, but also by what we 
assume about the presently unmodelled relation between the average densities inside the core radius and the 
radius of efficient self-interactions, as parameterized by $\kappa$ defined in Eq.~(\ref{eq:kappa}).
For the range of cross sections that we are interested in here, we find in fact 
that the bound on $\tilde\sigma/m_\chi$ scales roughly linearly with the 
maximal value of $\kappa$  allowed in our analysis.
While data from simulations
already strongly constrain the (average) value of $\kappa$ to be much larger than what we have 
adopted in our analysis, more simulation data is thus needed in order to make this bound
more robust.

We note that the translation to a bound on the {\it physical} cross section adds, as 
stated in arriving at Eq.~(\ref{eq:smax}), another uncertainty
to the present analysis -- though one should stress that this uncertainty is shared by most analyses 
constraining SIDM (which is rarely spelled out explicitly).
In other words, this does not affect the 
conclusion that our method  to compare theory with observations is more robust than those based 
on individual objects. 
 In any case, we clearly expect this to be an $\mathcal{O}(1)$ 
effect, at most, something which we have checked explicitly by comparison to a limited set of 
simulations (see Fig.~\ref{fig:rochasigma} and, for a similar test, Ref.~\cite{Kaplinghat:2015aga}),
and which we think is captured in the approximate limit stated in Eq.~(\ref{eq:stildemax}).
Still, we caution that in order to make the limit on the physical cross section more robust
would require a further refinement of the analytical model, i.e.~the formula (\ref{sidm_ansatz}),
which has been used in a very similar way previously in the literature.
This, in turn, calls for a more systematic study of the properties of simulated halos, which is beyond the
scope of this article.

Another potential worry might be that our bound is mostly driven by a small subset of objects -- which 
one then could argue may suffer from large systematic uncertainties, similar to bounds derived from 
individual objects. The dashed and dotted lines in Fig.~\ref{fig:like} essentially demonstrate that this
is {\it not} the case: both when excluding clusters (dashed line) and when excluding (all other) objects that 
are baryon-dominated in their central parts (dotted line) from our analysis, the limit does not change by  
more than what can be explained by the reduced sample size.
Let us point out that these two model classes are indeed the main suspects when looking for such an effect:
\begin{itemize}
\item Observations of colliding {\it clusters} lead to the strongest currently existing constraints on SIDM,
with $\sigma/m_\chi\lesssim 0.47$\,cm$^2$/g, and it has been argued that their small cores (if any) 
lead to even stronger bounds \cite{Kaplinghat:2015aga}. While not undisputed,
this has triggered much phenomenological interest in velocity-dependent self-interactions in order to evade cluster bounds and at the same time allow 
for $\sigma/m_\chi\sim 1$\,cm$^2$/g  at (dwarf) 
galaxy scales, where the typical DM velocities are up to one order of magnitude smaller 
\cite{Yoshida:2000uw,Firmani:2000ce,Firmani:2000qe,Colin:2002nk} (and more 
recently~\cite{Loeb:2010gj,Aarssen:2012fx,Kaplinghat:2015aga}). 
Our (cluster) bounds, hence, do not show a significant velocity dependence.
\item Large {\it baryon} densities in the inner halo parts can lead to contracted profiles, instead of cores,
counteracting the effect of core-formation due to self-interactions \cite{Kamada:2016euw}. Such an effect could ``hide''
large SIDM cross sections, and hence potentially spoil our claim of deriving upper bounds on SIDM 
irrespective of the role of baryons. While we on purpose did not include any objects
that are close to being as baryon-dominated as the corresponding examples in Ref.~\cite{Kamada:2016euw}, a remaining worry
might be that a similar effect could be seen in the small number of objects in our sample where the 
impact of baryons on the gravitational potential in the {\it inner} parts of the halo is similar to that of DM. 
As the dotted line shows, this is not a concern as removing those objects from our analysis does not significantly
weaken the bound on $\tilde\sigma/m_\chi$. 
\end{itemize}
We also checked, independently, the constraining power of dSphs alone (dash-dotted line in Fig.~\ref{fig:like}).
This much weaker bound ($\tilde \sigma/m_\chi \lesssim 1.4$\,cm$^2$/g) is mostly driven by 
Sculptor and Draco, leading respectively to $\tilde\sigma/m_\chi \leq 3.5$\,cm$^2$/g and 
$\sigma/m_\chi \leq 4.4$\,cm$^2$/g, while Carina alone formally allows for a cross section of 
$\tilde\sigma/m_\chi\sim50$\,cm$^2$/g. While this appears broadly consistent with what was  found in 
Ref.~\cite{Valli:2017ktb}, we stress that one cannot easily compare these results as we do not impose
a cosmological prior on the $c-M_{\rm 200}$ relation (or the distribution of circular velocities). 
From the discussion in Section \ref{sec:cores}, in fact, we would expect  that kinematic data from classical 
dSphs alone would lead to even much weaker constraints (see for example Fig.~\ref{fig:Carina}).
The solution to this apparent paradox is that, as already stressed above, also in our analysis 
it is not the kinematic data alone that set the constraints. For example, allowing $\kappa$ to be as large as 
20 -- which is far larger than anything found in simulations -- would imply that our dSphs bound
relaxes from $\tilde \sigma/m_\chi \lesssim 1.4$\,cm$^2$/g to $\tilde \sigma/m_\chi \lesssim 4.2$\,cm$^2$/g. In 
general, we find again that the bound scales linearly with the maximally allowed value for $\kappa$.

In this article, we have presented
a new method to test SIDM and, as a proof of concept, derived stringent constraints already from a relatively
small sample of objects. Prospects to derive more stringent limits, simply by increasing
the sample size, are thus obviously promising. This would be  particularly interesting for dwarf-scale 
field halos, for which currently no fits to cored profiles exist 
An even more promising way to significantly improve
our limits is to reduce the errors on the observational parameters that enter our analysis, i.e.~the
experimental core radius and the average density inside this radius. This can be achieved if (the product of)
these quantities is {\it directly} constrained in the kinematic analysis, rather than taking the detour
via first fitting a density profile to the data. Recalling that the surface density is also very useful for
understanding basic scaling laws of $\Lambda$CDM cosmology, viz.~the concentration-mass relation, 
we use the opportunity for a general ``plea'' to observers: for many applications, {\it it is more advantageous
to present measurements of the average (surface) density of DM halos than fits to given profile parameterizations}.

\section{Conclusions}
\label{sec:conc}

Self-interacting dark matter (SIDM) has been the subject of increasing interest in the last 
few years, both because it may provide a solution to the long-standing small-scale problems
of the cosmological concordance model, and because it opens up interesting avenues 
for model-building that involve DM particles which could not be detected by traditional 
means. In this article we have introduced a new method to constrain such DM self-interactions
by re-visiting the surface density of astronomical objects ranging from dwarf galaxies to 
galaxy clusters. The main advantage of our method is that it is based on {\it ensembles} 
of objects, implying that the resulting constraints are rather robust and less sensitive to 
the often poorly understood astrophysical properties of {\it individual} objects.
The other crucial input to our analysis is how the average density inside
the region of efficient self-interactions is related to the core density; we obtain this
ratio from a direct comparison to simulations.

We illustrated our method by selecting a sample of around 50 objects, as described in 
App.~\ref{app:data}, where both cored and NFW profiles have been fitted to the available
kinematic data. We inferred the surface density at the experimental core radius for these
objects, and compared it to the surface density 
expected for an SIDM scenario 
with given cross section, cf.~Eq.~(\ref{eq:SDthnew}) and Fig.~\ref{fig:theoryexp2}. This
allowed us to construct a total likelihood, containing all halo objects, as a function of the 
self-scattering cross section (Section \ref{sec:stat}) and derive an upper limit of 
approximately 
$\sim0.3 \,{\rm cm}^2/{\rm g}$ on the cross section per unit mass,  $\sigma/m_\chi$ 
(Section \ref{sec:res}).
There are two main uncertainties entering this result: 
{\it i)} the limit on the {\it effective} cross section stated in Eq.~(\ref{eq:stildemax}) is driven to 
a large extent by the ratio of average densities as introduced in Eq.~(\ref{eq:kappa}), which
we currently constrain only from a relatively small number of simulations; {\it ii)} translating this 
to a bound on the  {\it physical} cross section involves an inevitable uncertainty 
in our theoretical prescription of the effect of SIDM, which we currently capture in the effective
parameter $\xi$ in (the commonly adopted) Eq.~(\ref{sidm_ansatz}). While we believe that our
quoted final bound does encapsulate this uncertainty, a more systematic investigation
of the properties of simulated SIDM halos -- e.g.~by considering larger samples than 
what is shown in Fig.~\ref{fig:rochasigma} -- has the clear potential of resulting in 
even stronger and more robust bounds.
As a byproduct of our analysis, we 
also revisited the scaling relation for the surface density evaluated further away from the center, 
at the scale radius of the NFW profile, and derived an updated version of the standard 
$c$-$M_{200}$ relation of $\Lambda$CDM cosmology, c.f.~Eq.~(\ref{our_cM}) and 
Fig.~\ref{fig:SDNFW}.

We note that for a velocity-independent self-interaction strength our upper bound is formally 
below, but still relatively close to, the $\sim1\,{\rm cm}^2/{\rm g}$ that have been reported 
as a requirement to fully address the $\Lambda$CDM small-scale problems without invoking 
baryonic explanations. 
On the other hand, our bound can currently not (yet) constrain this idea in scenarios where 
the self-interaction cross section is velocity-dependent and drops significantly below $1\,{\rm cm}^2/{\rm g}$ 
for masses larger than those of dwarf galaxies.
While still subject to some uncertainties, the reason for the relatively strong 
bound we report here is a combination of  the large number of objects that we include 
in our analysis and the fact that the parameter $\kappa$ introduced in Eq.~(\ref{eq:kappa}) is so 
well constrained by simulations.
As we have discussed in more detail in Section \ref{sec:disc}, furthermore, prospects for 
future significant improvements of these bounds are very good. This may finally settle the 
question of whether DM has astrophysically
relevant self-interactions, thereby providing yet another example of how useful observational
scaling laws are in astronomy.

\bigskip
\acknowledgments
We thank Carlos Frenk, Manoj Kaplinghat, Matthew Walker, Mauro Valli and Jes\'us Zavala
for very valuable communications during the preparation of this work, as well as all 
participants of the workshop on 
``self-interacting dark matter'' in Copenhagen (31 July - 4 August) for stimulating discussions.


\newpage
\appendix
\section{Datasets}
\label{app:data}

For our analysis, we select objects where the density profile is reasonably 
well fit to both a cored profile and an NFW profile at large radii, and 
where the two resulting masses inside the best measured distance are consistent 
with each other. 
The cored profiles that we take into account are:
\begin{itemize}
    \item 
    Burkert profile
    \begin{equation}
    \rho_{\text{Burk}}(r) = 
    \frac{\rho_0}{(1 + r/r_0) (1 + (r/r_0)^2)},
    \label{eq:Burkert}
    \end{equation}
    \item Pseudoisothermal profile
    \begin{equation}
    \rho_{\text{ISO}}(r) = 
    \frac{\rho_0}{ (1 + (r/r_0)^2)},
    \end{equation}
    \item Modified pseudoisothermal profile
    \begin{equation}
    \rho_{\text{ISO2}}(r) = 
    \frac{\rho_0}{ (1 + (r/r_0)^2)^{3/2}},
    \label{eq:ISO2}
    \end{equation}
    \item Cored NFW profile:
    \begin{equation}
    \rho_{\text{cNFW}}=\frac{b \rho_s}{(1 + b r/r_s)(1 + r/r_s)^2}.
    \end{equation} 
\end{itemize}

The presence of baryons can significantly modify the ``isothermal'' solutions of the Jeans equation
 that feature $\sigma_v\approx const.$ for $r<r_\mathrm{SIDM}$, up to the  point where instead of cored 
 profiles one finds cuspy profiles \cite{Kamada:2016euw}. These extreme cases are obviously not
among the objects that we consider, given that we demand a good fit to a cored profile.
 In order to keep the discussion of core radius vs.~self-interaction radius as model-independent as possible, 
however, we would still like to make sure that we only select objects that are sufficiently DM dominated for
the baryons not to affect our analysis, e.g.~via poorly known mass-to-light ratios for stars. 
Concretely, we demand that at the largest measured radius the mass of the DM is at least 4 times 
larger than the mass of the baryons. This implies that we keep {\it some} objects that have a larger baryon
content at the smallest observationally accessible distances; in Section \ref{sec:disc}, we discuss explicitly
the (small) impact of those objects on our final results.

The following is a complete list of objects that we selected for our analysis, followed by a brief description of
the characteristics of each object class: 

\begin{itemize}
    \item \textbf{Dwarf Spheroidal galaxies}, \cite{Salucci:2011ee,walker_private} (NFW and Burkert profiles): Sculptor, Leo I, Carina, Draco, Ursa Minor, Fornax
    \item \textbf{Low Surface Brightness}, \cite{KuziodeNaray:2007qi} (NFW and ISO profiles): F563-V2, F563-1, F583-4
    \item \textbf{GHASP}, \cite{Spano:2007nt} (NFW and ISO2): UGC-3876, UGC-4256, UGC-4456, UGC-4499, UGC-10310
    \item \textbf{THINGS}, \cite{deBlok:2008wp} (NFW and ISO profiles): NGC-2903 (outer), NGC-3198 (2 comp),NGC-2403 (1 comp),  NGC-2403 (2 comp),  NGC-2841, NGC-3621, DDO-154
    \item \textbf{LITTLE THINGS}, \cite{Oh:2015xoa} (NFW and ISO profiles): DDO-52, DDO-101, WLM, Haro-29, DDO-87, DDO-126, DDO-216
    \item \textbf{SPARC},~\cite{Lelli:2016zqa} (NFW and Burk): F568-V1, NGC-24, NGC-2683, NGC-3769, NGC-3953, NGC-3992, NGC-4100, NGC-4183, UGC-2259, UGC-5721, UGC-7690, UGC-8490, UGC-9992, UGC-12506
    \item \textbf{Clusters}~\cite{Newman:2012nw} (NFW and cNFW):  MS2137, A963, A383, A611, A2537, A2667
\end{itemize}

\paragraph{Dwarf Spheroidal galaxies}

\begin{figure}[t!]
  \centering
    \includegraphics[width=\textwidth, trim ={0 3.5cm 1cm 1cm}, clip]{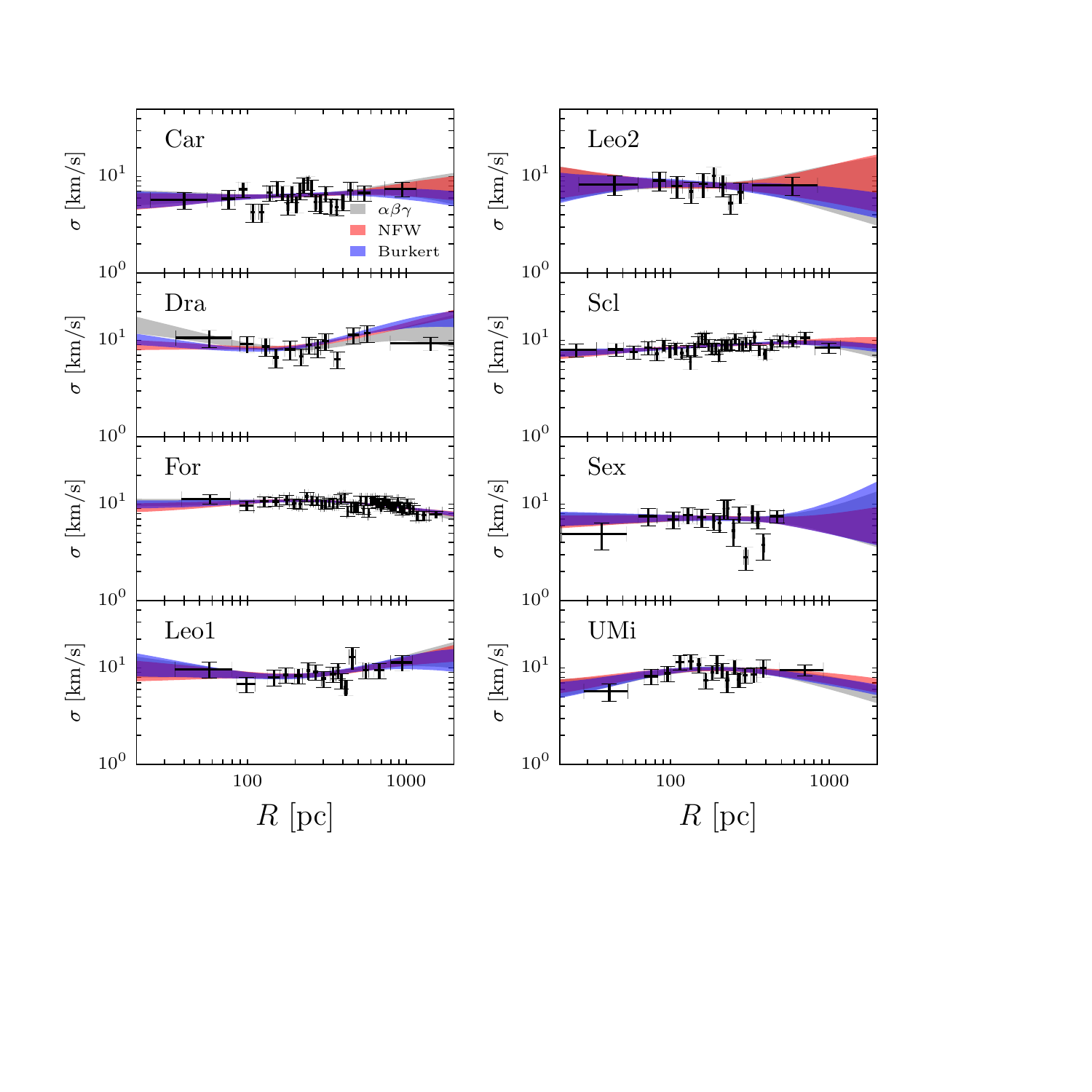}
    \caption{The velocity dispersion as a function of the galactocentric radius for  classical dSphs. Black points 
    correspond to observational data, and the red (blue, grey) line is the result of a fitting an NFW (Burkert, 
    generalized NFW) profile, with the bands indicating 1$\sigma$ uncertainty. The data from the generalized 
    NFW profile are taken from Ref.~\cite{Geringer-Sameth:2014yza}, while the NFW and Burkert fits are newly 
    calculated by M. G. Walker~\cite{walker_private}. 
    }
    \label{fig:dwarfsveldispersion}
\end{figure}

\begin{figure}[t!]
  \centering
    \includegraphics[width=\textwidth, trim ={0 3.5cm 1cm 1cm}, clip]{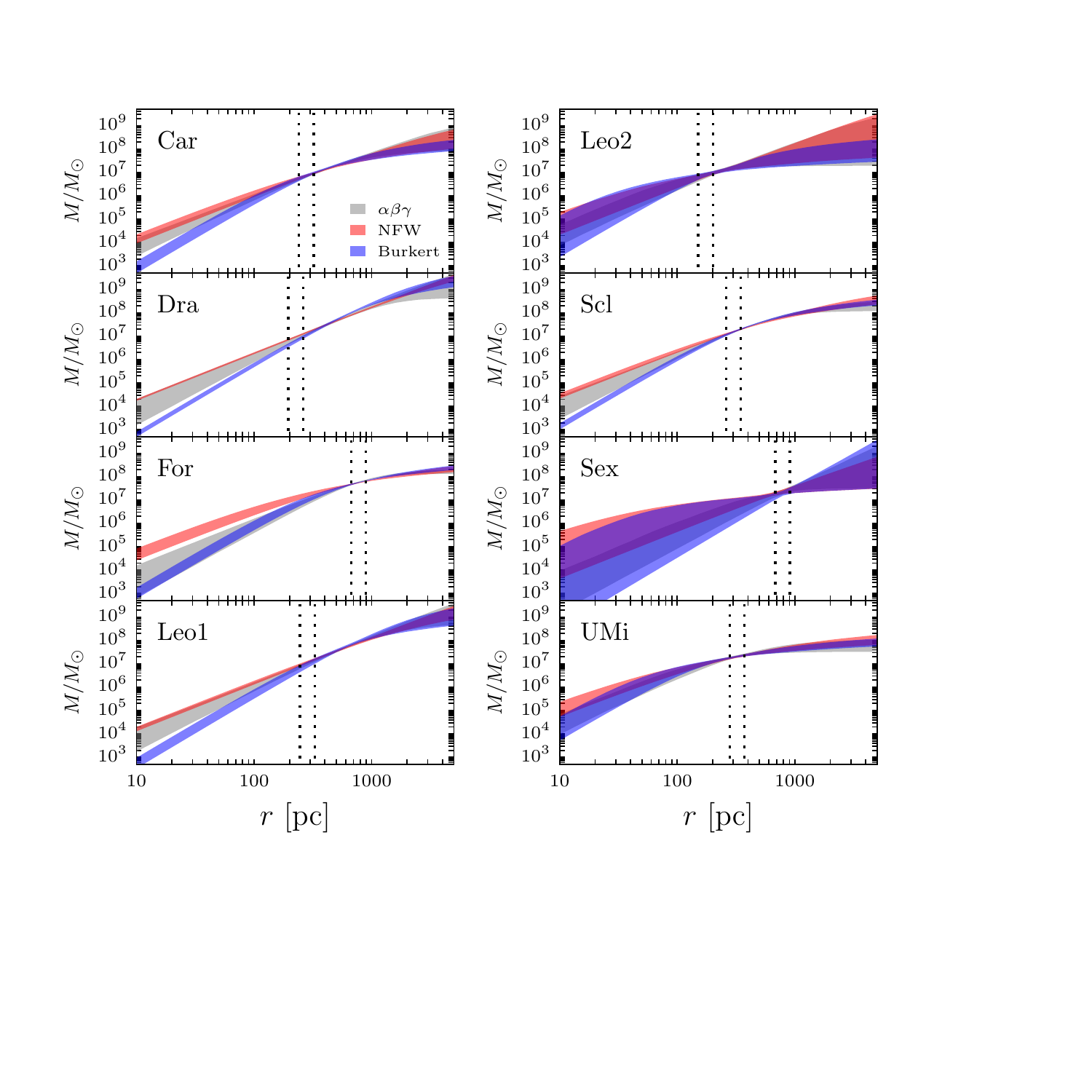}
    \caption{Same as Fig.~\ref{fig:dwarfsveldispersion}, but now for the mass profile. The dotted vertical lines show the position and uncertainty of the half-light radius $r_h$. The data from the generalized 
    NFW profile are taken from Ref.~\cite{Geringer-Sameth:2014yza}, while the NFW and Burkert fits are newly 
    calculated by M. G. Walker~\cite{walker_private}. 
    }
    \label{fig:dwarfsmassprofile}
\end{figure}

Dwarf spheroidal galaxies (dSphs) are satellites of the Milky Way and nearby galaxies. They are the objects with the 
highest known mass-to-light ratio and DM dominated even in the central parts, thus very suitable to study 
DM properties.
There are so-called classical dSphs with a relatively large amount of stars, such that their DM profile 
can be reconstructed sufficiently well, see Fig.~\ref{fig:dwarfsveldispersion} for velocity dispersion profiles 
and Fig.~\ref{fig:dwarfsmassprofile} with the reconstructed mass profiles. 
To minimize the uncertainty in our analysis, we will only consider  classical dSphs.

The observable quantities in dSphs are the velocity dispersion along the line of sight $\sigma_{\text{los}}$ and the 
half-light radius $r_h$. The main uncertainty on the mass distribution comes from the fact that one instead needs 
the full 3D velocity dispersion $\sigma$ to calculate the mass inside some radius $r$. 
One of the best mass estimates is obtained for the mass inside the half-light 
radius~\cite{Walker:2011zu,Wolf:2009tu,Amorisco:2011hb}, 
as clearly illustrated in Fig.~\ref{fig:dwarfsmassprofile}. To measure the slope of the central part of the DM profile 
it was proposed to generalize this idea 
to 2 or 3 distinct star populations, which led to the (still being disputed) claim of evidence for 
cores~\cite{Amorisco:2010ns}. 

\paragraph{Spiral and irregular galaxies}

In spiral galaxies, the measurable quantities are rotation curves for stars and neutral hydrogen. As these objects 
are dominated by  baryons in their central parts, the main uncertainty on the DM profile in this region comes 
from modelling the baryon mass. As a result, the DM profile can often be fitted (almost) equally well with NFW 
and cored profiles, 
see e.g.~the rotation curves displayed in Ref.~\cite{deBlok:2008wp}. The best measurement of 
the DM profile derives instead from the flat part of the rotation curve, where 
DM dominates over baryons.

Low surface brightness galaxies (LSBs) are diffuse galaxies with a surface brightness that, when viewed from 
Earth, is at least one magnitude lower than the ambient night sky. Most LSBs are dwarf galaxies, and most of their 
baryonic matter is in the form of neutral hydrogen gas, rather than stars. They appear to have over 95\% of their 
mass in DM, and are DM dominated even in the central parts. As in the case of spiral galaxies, the observables are 
rotation curves $v(r)$ of stars and neutral hydrogen. Reconstructed DM profiles appear to be more consistent 
with cored profiles~\cite{deBlok:2001hbg}.

\paragraph{Galaxy clusters}
The exceptional class of objects where we allow a larger baryon contribution in the inner parts are clusters of 
galaxies. This is because all clusters are baryon-dominated in the center by the existence of the brightest cluster 
galaxy and gas. On the other hand, it was claimed that some clusters show evidence
for cores in the DM profile \cite{Newman:2012nw}. Such an evidence for  
cores in clusters is highly disputed, see e.g.~the discussion in Ref.~\cite{Schaller:2014gwa}. For the current work 
we simply take the cored fits of the DM profiles from Ref.~\cite{Newman:2012nw}, without any additional 
selection of the objects (but demonstrate in Section \ref{sec:disc} that these objects only have a very minor
impact on our results).

\section{Halo age}
\label{app:age}

In this Appendix, we briefly describe how we implement the average halo age that 
enters the definition of the scale of self-interactions in Eq.~(\ref{sidm_ansatz}).
The ``formation redshift'' (the half-mass formation time) of a halo with present mass $M$ is 
implicitly given in terms of the critical overdensity, $\delta_{\rm crit} = (\rho-\rho_c)/\rho_c$,
by~\cite{Ludlow:2013vxa}
\begin{equation}
    \delta_{\text{crit}}(z_f) = \delta_{\text{crit}}^0/D(z_f) = \delta_{\text{crit}}^0 + 
    0.477 \sqrt{2 [\sigma^2(fM,0)-\sigma^2(M,0)]}\,.
    \label{zfdef}
\end{equation}
Here, the  critical density threshold for spherical collapse at $z = 0$ is given by
$\delta_{\text{crit}}^0=0.15(12\pi)^{2/3}\Omega_m^{0.055}$, and the linear 
growth factor by \cite{Ludlow:2016ifl}
\begin{equation}
D(z)=\frac{\Omega_M(z)}{\Omega_M^0} \frac{\Psi(0)}{\Psi(z)} (1+z)^{-1},
\end{equation}
with 
\begin{equation}
\Psi(z) = \Omega_M(z)^{4/7} - \Omega_{\Lambda}(z) + 
\left( 1+\frac{\Omega_M(z)}{2} \right) \left( 1+\frac{\Omega_{\Lambda}(z)}{70} \right)
\end{equation}
and
\begin{equation}
\Omega_{\Lambda}(z) = \frac{\Omega_{\Lambda}^0}{\Omega_{\Lambda}^0 + \Omega_{m}^0 (1+z)^3},
\end{equation}
These expressions assume a flat universe, $\Omega_m(z) = 1 - \Omega_{\Lambda}(z)$, and we take
the current value of the matter density $\Omega_m=\Omega_m^0=\rho_m^0/\rho_c^0$ 
as measured by Planck \cite{Ade:2015xua}.

The second step in Eq.~(\ref{zfdef}) involves a fitting parameter from the accretion histories, $f=0.068$,
and the linear rms fluctuation in spheres of mass $M$. The latter is given by \cite{Ludlow:2016ifl}
\begin{equation}
\sigma(M,z) = D(z) \frac{22.26 \xi^{0.292}}{1 + 1.53 \xi^{0.275} + 3.38 \xi^{0.198}},
\end{equation}
where
\begin{equation}
\xi = \left( \frac{M}{10^{10} h^{-1} M_{\odot}} \right)^{-1}.
\end{equation}

\begin{figure}[t]
  \centering
  \includegraphics[width=0.6\textwidth]{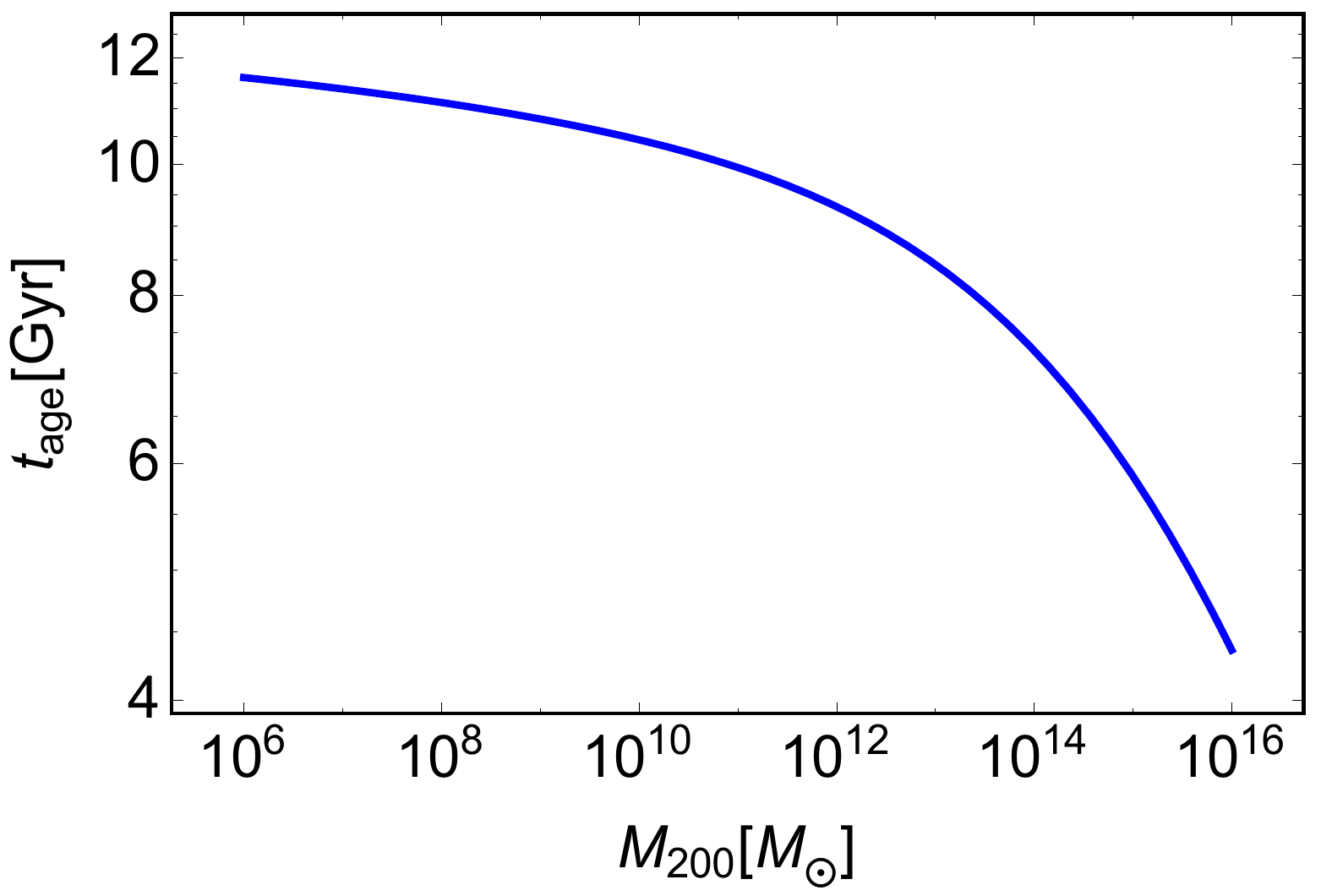}
  \caption{{Halo age as a function of the virial mass adopted from Refs.~\cite{Ludlow:2013vxa,Ludlow:2016ifl,Liu:2012kd}.}}
  \label{fig:tageofM200}
\end{figure}

Solving Eq.~(\ref{zfdef}) for the formation redshift, one finally gets the halo age as \cite{Liu:2012kd}
\begin{equation}
    t_{\text{age}}=t_0 - t(z_f),
\end{equation}
where $t_0$ is the current age of the Universe and
\begin{equation}
    t(z_f) = \frac{1}{H_0} \int\limits_{z_f}^{\infty}
    \frac{dz}{(1+z) \sqrt{\Omega_M (1+z)^3 + \Omega_{\Lambda}}}\,.
\end{equation}
We plot this function in Fig.~\ref{fig:tageofM200}.

\bibliographystyle{JHEP}
\bibliography{SIDM}

\providecommand{\href}[2]{#2}\begingroup\raggedright\begin{thebibliography}{10}

\bibitem{Ade:2015xua}
{\scshape Planck} collaboration, P.~A.~R. Ade et~al., \emph{{Planck 2015
  results. XIII. Cosmological parameters}},
  \href{https://doi.org/10.1051/0004-6361/201525830}{\emph{Astron. Astrophys.}
  {\bfseries 594} (2016) A13},
  [\href{https://arxiv.org/abs/1502.01589}{{\ttfamily 1502.01589}}].

\bibitem{Springel:2005mi}
V.~Springel, \emph{{The Cosmological simulation code GADGET-2}},
  \href{https://doi.org/10.1111/j.1365-2966.2005.09655.x}{\emph{Mon. Not. Roy.
  Astron. Soc.} {\bfseries 364} (2005) 1105--1134},
  [\href{https://arxiv.org/abs/astro-ph/0505010}{{\ttfamily
  astro-ph/0505010}}].

\bibitem{Vogelsberger:2014kha}
M.~Vogelsberger, S.~Genel, V.~Springel, P.~Torrey, D.~Sijacki, D.~Xu et~al.,
  \emph{{Properties of galaxies reproduced by a hydrodynamic simulation}},
  \href{https://doi.org/10.1038/nature13316}{\emph{Nature} {\bfseries 509}
  (2014) 177--182}, [\href{https://arxiv.org/abs/1405.1418}{{\ttfamily
  1405.1418}}].

\bibitem{Tulin:2017ara}
S.~Tulin and H.-B. Yu, \emph{{Dark Matter Self-interactions and Small Scale
  Structure}},  \href{https://arxiv.org/abs/1705.02358}{{\ttfamily
  1705.02358}}.

\bibitem{Aprile:2017iyp}
{\scshape XENON} collaboration, E.~Aprile et~al., \emph{{First Dark Matter
  Search Results from the XENON1T Experiment}},
  \href{https://doi.org/10.1103/PhysRevLett.119.181301}{\emph{Phys. Rev. Lett.}
  {\bfseries 119} (2017) 181301},
  [\href{https://arxiv.org/abs/1705.06655}{{\ttfamily 1705.06655}}].

\bibitem{Cui:2017nnn}
{\scshape PandaX-II} collaboration, X.~Cui et~al., \emph{{Dark Matter Results
  From 54-Ton-Day Exposure of PandaX-II Experiment}},
  \href{https://doi.org/10.1103/PhysRevLett.119.181302}{\emph{Phys. Rev. Lett.}
  {\bfseries 119} (2017) 181302},
  [\href{https://arxiv.org/abs/1708.06917}{{\ttfamily 1708.06917}}].

\bibitem{Bringmann:2013vra}
T.~Bringmann, J.~Hasenkamp and J.~Kersten, \emph{{Tight bonds between sterile
  neutrinos and dark matter}},
  \href{https://doi.org/10.1088/1475-7516/2014/07/042}{\emph{JCAP} {\bfseries
  1407} (2014) 042}, [\href{https://arxiv.org/abs/1312.4947}{{\ttfamily
  1312.4947}}].

\bibitem{Kaplinghat:2015aga}
M.~Kaplinghat, S.~Tulin and H.-B. Yu, \emph{{Dark Matter Halos as Particle
  Colliders: Unified Solution to Small-Scale Structure Puzzles from Dwarfs to
  Clusters}}, \href{https://doi.org/10.1103/PhysRevLett.116.041302}{\emph{Phys.
  Rev. Lett.} {\bfseries 116} (2016) 041302},
  [\href{https://arxiv.org/abs/1508.03339}{{\ttfamily 1508.03339}}].

\bibitem{Valli:2017ktb}
M.~Valli and H.-B. Yu, \emph{{Dark matter self-interactions from the internal
  dynamics of dwarf spheroidals}},
  \href{https://arxiv.org/abs/1711.03502}{{\ttfamily 1711.03502}}.

\bibitem{Spergel:1999mh}
D.~N. Spergel and P.~J. Steinhardt, \emph{{Observational evidence for
  selfinteracting cold dark matter}},
  \href{https://doi.org/10.1103/PhysRevLett.84.3760}{\emph{Phys. Rev. Lett.}
  {\bfseries 84} (2000) 3760--3763},
  [\href{https://arxiv.org/abs/astro-ph/9909386}{{\ttfamily
  astro-ph/9909386}}].

\bibitem{Loeb:2010gj}
A.~Loeb and N.~Weiner, \emph{{Cores in Dwarf Galaxies from Dark Matter with a
  Yukawa Potential}},
  \href{https://doi.org/10.1103/PhysRevLett.106.171302}{\emph{Phys. Rev. Lett.}
  {\bfseries 106} (2011) 171302},
  [\href{https://arxiv.org/abs/1011.6374}{{\ttfamily 1011.6374}}].

\bibitem{Vogelsberger:2012ku}
M.~Vogelsberger, J.~Zavala and A.~Loeb, \emph{{Subhaloes in Self-Interacting
  Galactic Dark Matter Haloes}},
  \href{https://doi.org/10.1111/j.1365-2966.2012.21182.x}{\emph{Mon. Not. Roy.
  Astron. Soc.} {\bfseries 423} (2012) 3740},
  [\href{https://arxiv.org/abs/1201.5892}{{\ttfamily 1201.5892}}].

\bibitem{Peter:2012jh}
A.~H.~G. Peter, M.~Rocha, J.~S. Bullock and M.~Kaplinghat, \emph{{Cosmological
  Simulations with Self-Interacting Dark Matter II: Halo Shapes vs.
  Observations}}, \href{https://doi.org/10.1093/mnras/sts535}{\emph{Mon. Not.
  Roy. Astron. Soc.} {\bfseries 430} (2013) 105},
  [\href{https://arxiv.org/abs/1208.3026}{{\ttfamily 1208.3026}}].

\bibitem{Zavala:2012us}
J.~Zavala, M.~Vogelsberger and M.~G. Walker, \emph{{Constraining
  Self-Interacting Dark Matter with the Milky Way's dwarf spheroidals}},
  \href{https://doi.org/10.1093/mnrasl/sls053}{\emph{Mon. Not. Roy. Astron.
  Soc.} {\bfseries 431} (2013) L20--L24},
  [\href{https://arxiv.org/abs/1211.6426}{{\ttfamily 1211.6426}}].

\bibitem{Aarssen:2012fx}
L.~G. van~den Aarssen, T.~Bringmann and C.~Pfrommer, \emph{{Is dark matter with
  long-range interactions a solution to all small-scale problems of
  $\Lambda$CDM cosmology?}},
  \href{https://doi.org/10.1103/PhysRevLett.109.231301}{\emph{Phys. Rev. Lett.}
  {\bfseries 109} (2012) 231301},
  [\href{https://arxiv.org/abs/1205.5809}{{\ttfamily 1205.5809}}].

\bibitem{Elbert:2014bma}
O.~D. Elbert, J.~S. Bullock, S.~Garrison-Kimmel, M.~Rocha, J.~Oñorbe and
  A.~H.~G. Peter, \emph{{Core formation in dwarf haloes with self-interacting
  dark matter: no fine-tuning necessary}},
  \href{https://doi.org/10.1093/mnras/stv1470}{\emph{Mon. Not. Roy. Astron.
  Soc.} {\bfseries 453} (2015) 29--37},
  [\href{https://arxiv.org/abs/1412.1477}{{\ttfamily 1412.1477}}].

\bibitem{Kamada:2016euw}
A.~Kamada, M.~Kaplinghat, A.~B. Pace and H.-B. Yu, \emph{{How the
  Self-Interacting Dark Matter Model Explains the Diverse Galactic Rotation
  Curves}}, \href{https://doi.org/10.1103/PhysRevLett.119.111102}{\emph{Phys.
  Rev. Lett.} {\bfseries 119} (2017) 111102},
  [\href{https://arxiv.org/abs/1611.02716}{{\ttfamily 1611.02716}}].

\bibitem{Robertson:2017mgj}
A.~Robertson, R.~Massey, V.~Eke, S.~Tulin, H.-B. Yu, Y.~Bahé et~al.,
  \emph{{The diverse density profiles of galaxy clusters with self-interacting
  dark matter plus baryons}},
  \href{https://arxiv.org/abs/1711.09096}{{\ttfamily 1711.09096}}.

\bibitem{Bullock:2017xww}
J.~S. Bullock and M.~Boylan-Kolchin, \emph{{Small-Scale Challenges to the
  $\Lambda$CDM Paradigm}},
  \href{https://doi.org/10.1146/annurev-astro-091916-055313}{\emph{Ann. Rev.
  Astron. Astrophys.} {\bfseries 55} (2017) 343--387},
  [\href{https://arxiv.org/abs/1707.04256}{{\ttfamily 1707.04256}}].

\bibitem{Flores:1994gz}
R.~A. Flores and J.~R. Primack, \emph{{Observational and theoretical
  constraints on singular dark matter halos}},
  \href{https://doi.org/10.1086/187350}{\emph{Astrophys. J.} {\bfseries 427}
  (1994) L1--4}, [\href{https://arxiv.org/abs/astro-ph/9402004}{{\ttfamily
  astro-ph/9402004}}].

\bibitem{1994Natur.370..629M}
B.~{Moore}, \emph{{Evidence against dissipation-less dark matter from
  observations of galaxy haloes}},
  \href{https://doi.org/10.1038/370629a0}{\emph{Nature} {\bfseries 370} (Aug.,
  1994) 629--631}.

\bibitem{2011MNRAS.415L..40B}
M.~{Boylan-Kolchin}, J.~S. {Bullock} and M.~{Kaplinghat}, \emph{{Too big to
  fail? The puzzling darkness of massive Milky Way subhaloes}},
  \href{https://doi.org/10.1111/j.1745-3933.2011.01074.x}{\emph{Mon. Not. Roy.
  Astron. Soc.} {\bfseries 415} (July, 2011) L40--L44},
  [\href{https://arxiv.org/abs/1103.0007}{{\ttfamily 1103.0007}}].

\bibitem{BoylanKolchin:2011dk}
M.~Boylan-Kolchin, J.~S. Bullock and M.~Kaplinghat, \emph{{The Milky Way's
  bright satellites as an apparent failure of LCDM}},
  \href{https://doi.org/10.1111/j.1365-2966.2012.20695.x}{\emph{Mon. Not. Roy.
  Astron. Soc.} {\bfseries 422} (2012) 1203--1218},
  [\href{https://arxiv.org/abs/1111.2048}{{\ttfamily 1111.2048}}].

\bibitem{Oman:2015xda}
K.~A. Oman et~al., \emph{{The unexpected diversity of dwarf galaxy rotation
  curves}}, \href{https://doi.org/10.1093/mnras/stv1504}{\emph{Mon. Not. Roy.
  Astron. Soc.} {\bfseries 452} (2015) 3650--3665},
  [\href{https://arxiv.org/abs/1504.01437}{{\ttfamily 1504.01437}}].

\bibitem{Oman:2016zjn}
K.~A. Oman, J.~F. Navarro, L.~V. Sales, A.~Fattahi, C.~S. Frenk, T.~Sawala
  et~al., \emph{{Missing dark matter in dwarf galaxies?}},
  \href{https://doi.org/10.1093/mnras/stw1251}{\emph{Mon. Not. Roy. Astron.
  Soc.} {\bfseries 460} (2016) 3610--3623},
  [\href{https://arxiv.org/abs/1601.01026}{{\ttfamily 1601.01026}}].

\bibitem{Moore:1999nt}
B.~Moore, S.~Ghigna, F.~Governato, G.~Lake, T.~R. Quinn, J.~Stadel et~al.,
  \emph{{Dark matter substructure within galactic halos}},
  \href{https://doi.org/10.1086/312287}{\emph{Astrophys. J.} {\bfseries 524}
  (1999) L19--L22}, [\href{https://arxiv.org/abs/astro-ph/9907411}{{\ttfamily
  astro-ph/9907411}}].

\bibitem{Klypin:1999uc}
A.~A. Klypin, A.~V. Kravtsov, O.~Valenzuela and F.~Prada, \emph{{Where are the
  missing Galactic satellites?}},
  \href{https://doi.org/10.1086/307643}{\emph{Astrophys. J.} {\bfseries 522}
  (1999) 82--92}, [\href{https://arxiv.org/abs/astro-ph/9901240}{{\ttfamily
  astro-ph/9901240}}].

\bibitem{Fattahi:2016nld}
A.~Fattahi, J.~F. Navarro, T.~Sawala, C.~S. Frenk, L.~V. Sales, K.~Oman et~al.,
  \emph{{The cold dark matter content of Galactic dwarf spheroidals: no cores,
  no failures, no problem}},
  \href{https://arxiv.org/abs/1607.06479}{{\ttfamily 1607.06479}}.

\bibitem{Mashchenko:2007jp}
S.~Mashchenko, J.~Wadsley and H.~M.~P. Couchman, \emph{{Stellar Feedback in
  Dwarf Galaxy Formation}},
  \href{https://doi.org/10.1126/science.1148666}{\emph{Science} {\bfseries 319}
  (2008) 174}, [\href{https://arxiv.org/abs/0711.4803}{{\ttfamily 0711.4803}}].

\bibitem{2012MNRAS.424.2715W}
J.~{Wang}, C.~S. {Frenk}, J.~F. {Navarro}, L.~{Gao} and T.~{Sawala}, \emph{{The
  missing massive satellites of the Milky Way}},
  \href{https://doi.org/10.1111/j.1365-2966.2012.21357.x}{\emph{Mon. Not. Roy.
  Astron. Soc.} {\bfseries 424} (Aug., 2012) 2715--2721},
  [\href{https://arxiv.org/abs/1203.4097}{{\ttfamily 1203.4097}}].

\bibitem{2015MNRAS.454..550G}
Q.~{Guo}, A.~P. {Cooper}, C.~{Frenk}, J.~{Helly} and W.~A. {Hellwing},
  \emph{{The Milky Way system in {$\Lambda$} cold dark matter cosmological
  simulations}}, \href{https://doi.org/10.1093/mnras/stv1938}{\emph{Mon. Not.
  Roy. Astron. Soc.} {\bfseries 454} (Nov., 2015) 550--559},
  [\href{https://arxiv.org/abs/1503.08508}{{\ttfamily 1503.08508}}].

\bibitem{Donato:2009ab}
F.~Donato, G.~Gentile, P.~Salucci, C.~F. Martins, M.~I. Wilkinson, G.~Gilmore
  et~al., \emph{{A constant dark matter halo surface density in galaxies}},
  \href{https://doi.org/10.1111/j.1365-2966.2009.15004.x}{\emph{Mon. Not. Roy.
  Astron. Soc.} {\bfseries 397} (2009) 1169--1176},
  [\href{https://arxiv.org/abs/0904.4054}{{\ttfamily 0904.4054}}].

\bibitem{Gentile:2009bw}
G.~Gentile, B.~Famaey, H.~Zhao and P.~Salucci, \emph{{Universality of galactic
  surface densities within one dark halo scale-length}},
  \href{https://doi.org/10.1038/nature08437}{\emph{Nature} {\bfseries 461}
  (2009) 627}, [\href{https://arxiv.org/abs/0909.5203}{{\ttfamily 0909.5203}}].

\bibitem{Boyarsky:2009rb}
A.~Boyarsky, O.~Ruchayskiy, D.~Iakubovskyi, A.~V. Maccio' and D.~Malyshev,
  \emph{{New evidence for dark matter}},
  \href{https://arxiv.org/abs/0911.1774}{{\ttfamily 0911.1774}}.

\bibitem{Boyarsky:2009af}
A.~Boyarsky, A.~Neronov, O.~Ruchayskiy and I.~Tkachev, \emph{{Universal
  properties of Dark Matter halos}},
  \href{https://doi.org/10.1103/PhysRevLett.104.191301}{\emph{Phys. Rev. Lett.}
  {\bfseries 104} (2010) 191301},
  [\href{https://arxiv.org/abs/0911.3396}{{\ttfamily 0911.3396}}].

\bibitem{Navarro:1995iw}
J.~F. Navarro, C.~S. Frenk and S.~D.~M. White, \emph{{The Structure of cold
  dark matter halos}}, \href{https://doi.org/10.1086/177173}{\emph{Astrophys.
  J.} {\bfseries 462} (1996) 563--575},
  [\href{https://arxiv.org/abs/astro-ph/9508025}{{\ttfamily
  astro-ph/9508025}}].

\bibitem{Navarro:1996gj}
J.~F. Navarro, C.~S. Frenk and S.~D.~M. White, \emph{{A Universal density
  profile from hierarchical clustering}},
  \href{https://doi.org/10.1086/304888}{\emph{Astrophys. J.} {\bfseries 490}
  (1997) 493--508}, [\href{https://arxiv.org/abs/astro-ph/9611107}{{\ttfamily
  astro-ph/9611107}}].

\bibitem{1965TrAlm...5...87E}
J.~{Einasto}, \emph{{On the Construction of a Composite Model for the Galaxy
  and on the Determination of the System of Galactic Parameters}}, {\emph{Trudy
  Astrofizicheskogo Instituta Alma-Ata} {\bfseries 5} (1965) 87--100}.

\bibitem{Blumenthal:1985qy}
G.~R. Blumenthal, S.~M. Faber, R.~Flores and J.~R. Primack, \emph{{Contraction
  of Dark Matter Galactic Halos Due to Baryonic Infall}},
  \href{https://doi.org/10.1086/163867}{\emph{Astrophys. J.} {\bfseries 301}
  (1986) 27}.

\bibitem{Gnedin:2004cx}
O.~Y. Gnedin, A.~V. Kravtsov, A.~A. Klypin and D.~Nagai, \emph{{Response of
  dark matter halos to condensation of baryons: Cosmological simulations and
  improved adiabatic contraction model}},
  \href{https://doi.org/10.1086/424914}{\emph{Astrophys. J.} {\bfseries 616}
  (2004) 16--26}, [\href{https://arxiv.org/abs/astro-ph/0406247}{{\ttfamily
  astro-ph/0406247}}].

\bibitem{Gustafsson:2006gr}
M.~Gustafsson, M.~Fairbairn and J.~Sommer-Larsen, \emph{{Baryonic Pinching of
  Galactic Dark Matter Haloes}},
  \href{https://doi.org/10.1103/PhysRevD.74.123522}{\emph{Phys. Rev.}
  {\bfseries D74} (2006) 123522},
  [\href{https://arxiv.org/abs/astro-ph/0608634}{{\ttfamily
  astro-ph/0608634}}].

\bibitem{Sawala:2015cdf}
T.~Sawala et~al., \emph{{The APOSTLE simulations: solutions to the Local
  Group's cosmic puzzles}},
  \href{https://doi.org/10.1093/mnras/stw145}{\emph{Mon. Not. Roy. Astron.
  Soc.} {\bfseries 457} (2016) 1931--1943},
  [\href{https://arxiv.org/abs/1511.01098}{{\ttfamily 1511.01098}}].

\bibitem{Schaye:2014tpa}
J.~Schaye et~al., \emph{{The EAGLE project: Simulating the evolution and
  assembly of galaxies and their environments}},
  \href{https://doi.org/10.1093/mnras/stu2058}{\emph{Mon. Not. Roy. Astron.
  Soc.} {\bfseries 446} (2015) 521--554},
  [\href{https://arxiv.org/abs/1407.7040}{{\ttfamily 1407.7040}}].

\bibitem{Schaller:2014uwa}
M.~Schaller, C.~S. Frenk, R.~G. Bower, T.~Theuns, A.~Jenkins, J.~Schaye et~al.,
  \emph{{Baryon effects on the internal structure of $\Lambda$CDM haloes in the
  EAGLE simulations}}, \href{https://doi.org/10.1093/mnras/stv1067}{\emph{Mon.
  Not. Roy. Astron. Soc.} {\bfseries 451} (2015) 1247--1267},
  [\href{https://arxiv.org/abs/1409.8617}{{\ttfamily 1409.8617}}].

\bibitem{Wang:2015jpa}
L.~Wang, A.~A. Dutton, G.~S. Stinson, A.~V. Macciò, C.~Penzo, X.~Kang et~al.,
  \emph{{NIHAO project – I. Reproducing the inefficiency of galaxy formation
  across cosmic time with a large sample of cosmological hydrodynamical
  simulations}}, \href{https://doi.org/10.1093/mnras/stv1937}{\emph{Mon. Not.
  Roy. Astron. Soc.} {\bfseries 454} (2015) 83--94},
  [\href{https://arxiv.org/abs/1503.04818}{{\ttfamily 1503.04818}}].

\bibitem{Tollet:2015gqa}
E.~Tollet et~al., \emph{{NIHAO – IV: core creation and destruction in dark
  matter density profiles across cosmic time}},
  \href{https://doi.org/10.1093/mnras/stv2856}{\emph{Mon. Not. Roy. Astron.
  Soc.} {\bfseries 456} (2016) 3542--3552},
  [\href{https://arxiv.org/abs/1507.03590}{{\ttfamily 1507.03590}}].

\bibitem{Rocha:2012jg}
M.~Rocha, A.~H.~G. Peter, J.~S. Bullock, M.~Kaplinghat, S.~Garrison-Kimmel,
  J.~Onorbe et~al., \emph{{Cosmological Simulations with Self-Interacting Dark
  Matter I: Constant Density Cores and Substructure}},
  \href{https://doi.org/10.1093/mnras/sts514}{\emph{Mon. Not. Roy. Astron.
  Soc.} {\bfseries 430} (2013) 81--104},
  [\href{https://arxiv.org/abs/1208.3025}{{\ttfamily 1208.3025}}].

\bibitem{1987gady.book.....B}
J.~{Binney} and S.~{Tremaine}, \emph{{Galactic dynamics}}.
\newblock Princeton University Press, 1987.

\bibitem{Pollack:2014rja}
J.~Pollack, D.~N. Spergel and P.~J. Steinhardt, \emph{{Supermassive Black Holes
  from Ultra-Strongly Self-Interacting Dark Matter}},
  \href{https://doi.org/10.1088/0004-637X/804/2/131}{\emph{Astrophys. J.}
  {\bfseries 804} (2015) 131},
  [\href{https://arxiv.org/abs/1501.00017}{{\ttfamily 1501.00017}}].

\bibitem{Walker:2009zp}
M.~G. Walker, M.~Mateo, E.~W. Olszewski, J.~Penarrubia, N.~W. Evans and
  G.~Gilmore, \emph{{A Universal Mass Profile for Dwarf Spheroidal Galaxies}},
  \href{https://doi.org/10.1088/0004-637X/704/2/1274,
  10.1088/0004-637X/710/1/886}{\emph{Astrophys. J.} {\bfseries 704} (2009)
  1274--1287}, [\href{https://arxiv.org/abs/0906.0341}{{\ttfamily 0906.0341}}].

\bibitem{Newman:2012nw}
A.~B. Newman, T.~Treu, R.~S. Ellis and D.~J. Sand, \emph{{The Density Profiles
  of Massive, Relaxed Galaxy Clusters: II. Separating Luminous and Dark Matter
  in Cluster Cores}},
  \href{https://doi.org/10.1088/0004-637X/765/1/25}{\emph{Astrophys. J.}
  {\bfseries 765} (2013) 25},
  [\href{https://arxiv.org/abs/1209.1392}{{\ttfamily 1209.1392}}].

\bibitem{Hofmann:2001bi}
S.~Hofmann, D.~J. Schwarz and H.~Stoecker, \emph{{Damping scales of neutralino
  cold dark matter}},
  \href{https://doi.org/10.1103/PhysRevD.64.083507}{\emph{Phys. Rev.}
  {\bfseries D64} (2001) 083507},
  [\href{https://arxiv.org/abs/astro-ph/0104173}{{\ttfamily
  astro-ph/0104173}}].

\bibitem{Ludlow:2013vxa}
A.~D. Ludlow, J.~F. Navarro, R.~E. Angulo, M.~Boylan-Kolchin, V.~Springel,
  C.~Frenk et~al., \emph{{The mass–concentration–redshift relation of cold
  dark matter haloes}}, \href{https://doi.org/10.1093/mnras/stu483}{\emph{Mon.
  Not. Roy. Astron. Soc.} {\bfseries 441} (2014) 378--388},
  [\href{https://arxiv.org/abs/1312.0945}{{\ttfamily 1312.0945}}].

\bibitem{Ludlow:2016ifl}
A.~D. Ludlow, S.~Bose, R.~E. Angulo, L.~Wang, W.~A. Hellwing, J.~F. Navarro
  et~al., \emph{{The mass–concentration–redshift relation of cold and warm
  dark matter haloes}}, \href{https://doi.org/10.1093/mnras/stw1046}{\emph{Mon.
  Not. Roy. Astron. Soc.} {\bfseries 460} (2016) 1214--1232},
  [\href{https://arxiv.org/abs/1601.02624}{{\ttfamily 1601.02624}}].

\bibitem{Liu:2012kd}
G.~Liu, Y.~Lu, X.~Chen, Y.~Zhao, W.~Du and X.~Meng, \emph{{The Age-Redshift
  Relation For Luminous Red Galaxies Obtained From the Full Spectrum Fitting
  and Its Cosmological Implications}},
  \href{https://doi.org/10.1088/0004-637X/758/2/107}{\emph{Astrophys. J.}
  {\bfseries 758} (2012) 107},
  [\href{https://arxiv.org/abs/1208.6502}{{\ttfamily 1208.6502}}].

\bibitem{Elbert:2016dbb}
O.~D. Elbert, J.~S. Bullock, M.~Kaplinghat, S.~Garrison-Kimmel, A.~S. Graus and
  M.~Rocha, \emph{{A Testable Conspiracy: Simulating Baryonic Effects on
  Self-Interacting Dark Matter Halos}},
  \href{https://arxiv.org/abs/1609.08626}{{\ttfamily 1609.08626}}.

\bibitem{Creasey:2016jaq}
P.~Creasey, O.~Sameie, L.~V. Sales, H.-B. Yu, M.~Vogelsberger and J.~Zavala,
  \emph{{Spreading out and staying sharp – creating diverse rotation curves
  via baryonic and self-interaction effects}},
  \href{https://doi.org/10.1093/mnras/stx522}{\emph{Mon. Not. Roy. Astron.
  Soc.} {\bfseries 468} (2017) 2283--2295},
  [\href{https://arxiv.org/abs/1612.03903}{{\ttfamily 1612.03903}}].

\bibitem{Burkert:1995yz}
A.~Burkert, \emph{{The Structure of dark matter halos in dwarf galaxies}},
  \href{https://doi.org/10.1086/309560}{\emph{IAU Symp.} {\bfseries 171} (1996)
  175}, [\href{https://arxiv.org/abs/astro-ph/9504041}{{\ttfamily
  astro-ph/9504041}}].

\bibitem{walker_private}
M.~G. Walker. private communication.

\bibitem{Cardone:2010jb}
V.~F. Cardone and C.~Tortora, \emph{{Dark matter scaling relations in
  intermediate z haloes}},
  \href{https://doi.org/10.1111/j.1365-2966.2010.17398.x}{\emph{Mon. Not. Roy.
  Astron. Soc.} {\bfseries 409} (2010) 1570},
  [\href{https://arxiv.org/abs/1007.3673}{{\ttfamily 1007.3673}}].

\bibitem{Napolitano:2010cq}
N.~R. Napolitano, A.~J. Romanowsky and C.~Tortora, \emph{{The central dark
  matter content of early-type galaxies: scaling relations and connections with
  star formation histories}},
  \href{https://doi.org/10.1111/j.1365-2966.2010.16710.x}{\emph{Mon. Not. Roy.
  Astron. Soc.} {\bfseries 405} (2010) 2351},
  [\href{https://arxiv.org/abs/1003.1716}{{\ttfamily 1003.1716}}].

\bibitem{DelPopolo:2012eb}
A.~Del~Popolo, V.~Cardone and G.~Belvedere, \emph{{Surface Density of dark
  matter haloes on galactic and cluster scales}},
  \href{https://arxiv.org/abs/1212.6797}{{\ttfamily 1212.6797}}.

\bibitem{Maccio:2006wpz}
A.~V. Maccio', A.~A. Dutton, F.~C. van~den Bosch, B.~Moore, D.~Potter and
  J.~Stadel, \emph{{Concentration, Spin and Shape of Dark Matter Haloes:
  Scatter and the Dependence on Mass and Environment}},
  \href{https://doi.org/10.1111/j.1365-2966.2007.11720.x}{\emph{Mon. Not. Roy.
  Astron. Soc.} {\bfseries 378} (2007) 55--71},
  [\href{https://arxiv.org/abs/astro-ph/0608157}{{\ttfamily
  astro-ph/0608157}}].

\bibitem{Neto:2007vq}
A.~F. Neto, L.~Gao, P.~Bett, S.~Cole, J.~F. Navarro, C.~S. Frenk et~al.,
  \emph{{The statistics of lambda CDM Halo Concentrations}},
  \href{https://doi.org/10.1111/j.1365-2966.2007.12381.x}{\emph{Mon. Not. Roy.
  Astron. Soc.} {\bfseries 381} (2007) 1450--1462},
  [\href{https://arxiv.org/abs/0706.2919}{{\ttfamily 0706.2919}}].

\bibitem{Maccio:2008pcd}
A.~V. Maccio', A.~A. Dutton and F.~C. v.~d. Bosch, \emph{{Concentration, Spin
  and Shape of Dark Matter Haloes as a Function of the Cosmological Model:
  WMAP1, WMAP3 and WMAP5 results}},
  \href{https://doi.org/10.1111/j.1365-2966.2008.14029.x}{\emph{Mon. Not. Roy.
  Astron. Soc.} {\bfseries 391} (2008) 1940--1954},
  [\href{https://arxiv.org/abs/0805.1926}{{\ttfamily 0805.1926}}].

\bibitem{MunozCuartas:2010ig}
J.~C. Munoz-Cuartas, A.~V. Maccio, S.~Gottlober and A.~A. Dutton, \emph{{The
  Redshift Evolution of LCDM Halo Parameters: Concentration, Spin, and Shape}},
  \href{https://doi.org/10.1111/j.1365-2966.2010.17704.x}{\emph{Mon. Not. Roy.
  Astron. Soc.} {\bfseries 411} (2011) 584},
  [\href{https://arxiv.org/abs/1007.0438}{{\ttfamily 1007.0438}}].

\bibitem{2012MNRAS.423.3018P}
F.~{Prada}, A.~A. {Klypin}, A.~J. {Cuesta}, J.~E. {Betancort-Rijo} and
  J.~{Primack}, \emph{{Halo concentrations in the standard {$\Lambda$} cold
  dark matter cosmology}},
  \href{https://doi.org/10.1111/j.1365-2966.2012.21007.x}{\emph{Mon. Not. Roy.
  Astron. Soc.} {\bfseries 423} (July, 2012) 3018--3030},
  [\href{https://arxiv.org/abs/1104.5130}{{\ttfamily 1104.5130}}].

\bibitem{Klypin:2014kpa}
A.~Klypin, G.~Yepes, S.~Gottlober, F.~Prada and S.~Hess, \emph{{MultiDark
  simulations: the story of dark matter halo concentrations and density
  profiles}}, \href{https://doi.org/10.1093/mnras/stw248}{\emph{Mon. Not. Roy.
  Astron. Soc.} {\bfseries 457} (2016) 4340--4359},
  [\href{https://arxiv.org/abs/1411.4001}{{\ttfamily 1411.4001}}].

\bibitem{Pilipenko:2017iae}
S.~V. Pilipenko, M.~A. Sánchez-Conde, F.~Prada and G.~Yepes, \emph{{Pushing
  down the low-mass halo concentration frontier with the Lomonosov cosmological
  simulations}}, \href{https://doi.org/10.1093/mnras/stx2319}{\emph{Mon. Not.
  Roy. Astron. Soc.} {\bfseries 472} (2017) 4918},
  [\href{https://arxiv.org/abs/1703.06012}{{\ttfamily 1703.06012}}].

\bibitem{Comerford:2007xb}
J.~M. Comerford and P.~Natarajan, \emph{{The Observed Concentration-Mass
  Relation for Galaxy Clusters}},
  \href{https://doi.org/10.1111/j.1365-2966.2007.11934.x}{\emph{Mon. Not. Roy.
  Astron. Soc.} {\bfseries 379} (2007) 190--200},
  [\href{https://arxiv.org/abs/astro-ph/0703126}{{\ttfamily
  astro-ph/0703126}}].

\bibitem{Sereno:2013aod}
M.~Sereno and G.~Covone, \emph{{The mass-concentration relation in massive
  galaxy clusters at redshift ~1}},
  \href{https://doi.org/10.1093/mnras/stt1086}{\emph{Mon. Not. Roy. Astron.
  Soc.} {\bfseries 434} (2013) 878},
  [\href{https://arxiv.org/abs/1306.6096}{{\ttfamily 1306.6096}}].

\bibitem{Merten:2014wna}
J.~Merten et~al., \emph{{CLASH: The Concentration-Mass Relation of Galaxy
  Clusters}}, \href{https://doi.org/10.1088/0004-637X/806/1/4}{\emph{Astrophys.
  J.} {\bfseries 806} (2015) 4},
  [\href{https://arxiv.org/abs/1404.1376}{{\ttfamily 1404.1376}}].

\bibitem{Du:2015dua}
W.~Du, Z.~Fan, H.~Shan, G.-B. Zhao, G.~Covone, L.~Fu et~al.,
  \emph{{Mass-concentration relation of clusters of galaxies from CFHTLenS}},
  \href{https://doi.org/10.1088/0004-637X/814/2/120}{\emph{Astrophys. J.}
  {\bfseries 814} (2015) 120},
  [\href{https://arxiv.org/abs/1510.08193}{{\ttfamily 1510.08193}}].

\bibitem{Groener:2015cxa}
A.~M. Groener, D.~M. Goldberg and M.~Sereno, \emph{{The galaxy cluster
  concentration–mass scaling relation}},
  \href{https://doi.org/10.1093/mnras/stv2341}{\emph{Mon. Not. Roy. Astron.
  Soc.} {\bfseries 455} (2016) 892--919},
  [\href{https://arxiv.org/abs/1510.01961}{{\ttfamily 1510.01961}}].

\bibitem{Amodeo:2016wtq}
S.~Amodeo, S.~Ettori, R.~Capasso and M.~Sereno, \emph{{The relation between
  mass and concentration in X-ray galaxy clusters at high redshift}},
  \href{https://doi.org/10.1051/0004-6361/201527630}{\emph{Astron. Astrophys.}
  {\bfseries 590} (2016) A126},
  [\href{https://arxiv.org/abs/1604.02163}{{\ttfamily 1604.02163}}].

\bibitem{Cibirka:2016nhw}
N.~Cibirka et~al., \emph{{CODEX weak lensing: concentration of galaxy clusters
  at z$\sim$0.5}}, \href{https://doi.org/10.1093/mnras/stx484}{\emph{Mon. Not.
  Roy. Astron. Soc.} {\bfseries 468} (2017) 1092--1116},
  [\href{https://arxiv.org/abs/1612.06871}{{\ttfamily 1612.06871}}].

\bibitem{Umetsu:2015baa}
K.~Umetsu, A.~Zitrin, D.~Gruen, J.~Merten, M.~Donahue and M.~Postman,
  \emph{{CLASH: Joint Analysis of Strong-Lensing, Weak-Lensing Shear and
  Magnification Data for 20 Galaxy Clusters}},
  \href{https://doi.org/10.3847/0004-637X/821/2/116}{\emph{Astrophys. J.}
  {\bfseries 821} (2016) 116},
  [\href{https://arxiv.org/abs/1507.04385}{{\ttfamily 1507.04385}}].

\bibitem{Lieu:2017xkq}
M.~Lieu, W.~M. Farr, M.~Betancourt, G.~P. Smith, M.~Sereno and I.~G. McCarthy,
  \emph{{Hierarchical inference of the relationship between Concentration and
  Mass in Galaxy Groups and Clusters}},
  \href{https://doi.org/10.1093/mnras/stx686}{\emph{Mon. Not. Roy. Astron.
  Soc.} {\bfseries 468} (2017) 4872--4886},
  [\href{https://arxiv.org/abs/1701.00478}{{\ttfamily 1701.00478}}].

\bibitem{Lin:2015fza}
H.~W. Lin and A.~Loeb, \emph{{Scaling Relations of Halo Cores for
  Self-Interacting Dark Matter}},
  \href{https://doi.org/10.1088/1475-7516/2016/03/009}{\emph{JCAP} {\bfseries
  1603} (2016) 009}, [\href{https://arxiv.org/abs/1506.05471}{{\ttfamily
  1506.05471}}].

\bibitem{Rolke:2004mj}
W.~A. Rolke, A.~M. Lopez and J.~Conrad, \emph{{Limits and confidence intervals
  in the presence of nuisance parameters}},
  \href{https://doi.org/10.1016/j.nima.2005.05.068}{\emph{Nucl. Instrum. Meth.}
  {\bfseries A551} (2005) 493--503},
  [\href{https://arxiv.org/abs/physics/0403059}{{\ttfamily physics/0403059}}].

\bibitem{carlos_private}
C.~Frenk. private communication.

\bibitem{Yoshida:2000uw}
N.~Yoshida, V.~Springel, S.~D.~M. White and G.~Tormen, \emph{{Weakly
  self-interacting dark matter and the structure of dark halos}},
  \href{https://doi.org/10.1086/317306}{\emph{Astrophys. J.} {\bfseries 544}
  (2000) L87--L90}, [\href{https://arxiv.org/abs/astro-ph/0006134}{{\ttfamily
  astro-ph/0006134}}].

\bibitem{Firmani:2000ce}
C.~Firmani, E.~D'Onghia, V.~Avila-Reese, G.~Chincarini and X.~Hernandez,
  \emph{{Evidence of self-interacting cold dark matter from galactic to galaxy
  cluster scales}},
  \href{https://doi.org/10.1046/j.1365-8711.2000.03555.x}{\emph{Mon. Not. Roy.
  Astron. Soc.} {\bfseries 315} (2000) L29},
  [\href{https://arxiv.org/abs/astro-ph/0002376}{{\ttfamily
  astro-ph/0002376}}].

\bibitem{Firmani:2000qe}
C.~Firmani, E.~D'Onghia, G.~Chincarini, X.~Hernandez and V.~Avila-Reese,
  \emph{{Constraints on dark matter physics from dwarf galaxies through galaxy
  cluster haloes}},
  \href{https://doi.org/10.1046/j.1365-8711.2001.04030.x}{\emph{Mon. Not. Roy.
  Astron. Soc.} {\bfseries 321} (2001) 713},
  [\href{https://arxiv.org/abs/astro-ph/0005001}{{\ttfamily
  astro-ph/0005001}}].

\bibitem{Colin:2002nk}
P.~Colin, V.~Avila-Reese, O.~Valenzuela and C.~Firmani, \emph{{Structure and
  subhalo population of halos in a selfinteracting dark matter cosmology}},
  \href{https://doi.org/10.1086/344259}{\emph{Astrophys. J.} {\bfseries 581}
  (2002) 777--793}, [\href{https://arxiv.org/abs/astro-ph/0205322}{{\ttfamily
  astro-ph/0205322}}].

\bibitem{Salucci:2011ee}
P.~Salucci, M.~I. Wilkinson, M.~G. Walker, G.~F. Gilmore, E.~K. Grebel, A.~Koch
  et~al., \emph{{Dwarf spheroidal galaxy kinematics and spiral galaxy scaling
  laws}}, \href{https://doi.org/10.1111/j.1365-2966.2011.20144.x}{\emph{Mon.
  Not. Roy. Astron. Soc.} {\bfseries 420} (2012) 2034},
  [\href{https://arxiv.org/abs/1111.1165}{{\ttfamily 1111.1165}}].

\bibitem{KuziodeNaray:2007qi}
R.~Kuzio~de Naray, S.~S. McGaugh and W.~J.~G. de~Blok, \emph{{Mass Models for
  Low Surface Brightness Galaxies with High Resolution Optical Velocity
  Fields}}, \href{https://doi.org/10.1086/527543}{\emph{Astrophys. J.}
  {\bfseries 676} (2008) 920--943},
  [\href{https://arxiv.org/abs/0712.0860}{{\ttfamily 0712.0860}}].

\bibitem{Spano:2007nt}
M.~Spano, M.~Marcelin, P.~Amram, C.~Carignan, B.~Epinat and O.~Hernandez,
  \emph{{GHASP: An H-alpha kinematic survey of spiral and irregular galaxies.
  5. Dark matter distribution in 36 nearby spiral galaxies}},
  \href{https://doi.org/10.1111/j.1365-2966.2007.12545.x}{\emph{Mon. Not. Roy.
  Astron. Soc.} {\bfseries 383} (2008) 297--316},
  [\href{https://arxiv.org/abs/0710.1345}{{\ttfamily 0710.1345}}].

\bibitem{deBlok:2008wp}
W.~J.~G. de~Blok, F.~Walter, E.~Brinks, C.~Trachternach, S.-H. Oh and R.~C.
  Kennicutt, Jr., \emph{{High-Resolution Rotation Curves and Galaxy Mass Models
  from THINGS}},
  \href{https://doi.org/10.1088/0004-6256/136/6/2648}{\emph{Astron. J.}
  {\bfseries 136} (2008) 2648--2719},
  [\href{https://arxiv.org/abs/0810.2100}{{\ttfamily 0810.2100}}].

\bibitem{Oh:2015xoa}
S.-H. Oh et~al., \emph{{High-resolution mass models of dwarf galaxies from
  LITTLE THINGS}},
  \href{https://doi.org/10.1088/0004-6256/149/6/180}{\emph{Astron. J.}
  {\bfseries 149} (2015) 180},
  [\href{https://arxiv.org/abs/1502.01281}{{\ttfamily 1502.01281}}].

\bibitem{Lelli:2016zqa}
F.~Lelli, S.~S. McGaugh and J.~M. Schombert, \emph{{SPARC: Mass Models for 175
  Disk Galaxies with Spitzer Photometry and Accurate Rotation Curves}},
  \href{https://doi.org/10.3847/0004-6256/152/6/157}{\emph{arXiv} (2016) },
  [\href{https://arxiv.org/abs/1606.09251}{{\ttfamily 1606.09251}}].

\bibitem{Geringer-Sameth:2014yza}
A.~Geringer-Sameth, S.~M. Koushiappas and M.~Walker, \emph{{Dwarf galaxy
  annihilation and decay emission profiles for dark matter experiments}},
  \href{https://doi.org/10.1088/0004-637X/801/2/74}{\emph{Astrophys. J.}
  {\bfseries 801} (2015) 74},
  [\href{https://arxiv.org/abs/1408.0002}{{\ttfamily 1408.0002}}].

\bibitem{Walker:2011zu}
M.~G. Walker and J.~Penarrubia, \emph{{A Method for Measuring (Slopes of) the
  Mass Profiles of Dwarf Spheroidal Galaxies}},
  \href{https://doi.org/10.1088/0004-637X/742/1/20}{\emph{Astrophys. J.}
  {\bfseries 742} (2011) 20},
  [\href{https://arxiv.org/abs/1108.2404}{{\ttfamily 1108.2404}}].

\bibitem{Wolf:2009tu}
J.~Wolf, G.~D. Martinez, J.~S. Bullock, M.~Kaplinghat, M.~Geha, R.~R. Munoz
  et~al., \emph{{Accurate Masses for Dispersion-supported Galaxies}},
  \href{https://doi.org/10.1111/j.1365-2966.2010.16753.x}{\emph{Mon. Not. Roy.
  Astron. Soc.} {\bfseries 406} (2010) 1220},
  [\href{https://arxiv.org/abs/0908.2995}{{\ttfamily 0908.2995}}].

\bibitem{Amorisco:2011hb}
N.~C. Amorisco and N.~W. Evans, \emph{{Dark Matter Cores and Cusps: The Case of
  Multiple Stellar Populations in Dwarf Spheroidals}},
  \href{https://doi.org/10.1111/j.1365-2966.2011.19684.x}{\emph{Mon. Not. Roy.
  Astron. Soc.} {\bfseries 419} (2012) 184--196},
  [\href{https://arxiv.org/abs/1106.1062}{{\ttfamily 1106.1062}}].

\bibitem{Amorisco:2010ns}
N.~C. Amorisco and N.~W. Evans, \emph{{Phase-space models of the dwarf
  spheroidals}},
  \href{https://doi.org/10.1111/j.1365-2966.2010.17715.x}{\emph{Mon. Not. Roy.
  Astron. Soc.} {\bfseries 411} (2011) 2118--2136},
  [\href{https://arxiv.org/abs/1009.1813}{{\ttfamily 1009.1813}}].

\bibitem{deBlok:2001hbg}
W.~J.~G. de~Blok, S.~S. McGaugh, A.~Bosma and V.~C. Rubin, \emph{{Mass density
  profiles of LSB galaxies}},
  \href{https://doi.org/10.1086/320262}{\emph{Astrophys. J.} {\bfseries 552}
  (2001) L23--L26}, [\href{https://arxiv.org/abs/astro-ph/0103102}{{\ttfamily
  astro-ph/0103102}}].

\bibitem{Schaller:2014gwa}
M.~Schaller, C.~S. Frenk, R.~G. Bower, T.~Theuns, J.~Trayford, R.~A. Crain
  et~al., \emph{{The effect of baryons on the inner density profiles of rich
  clusters}}, \href{https://doi.org/10.1093/mnras/stv1341}{\emph{Mon. Not. Roy.
  Astron. Soc.} {\bfseries 452} (2015) 343--355},
  [\href{https://arxiv.org/abs/1409.8297}{{\ttfamily 1409.8297}}].

\end{thebibliography}\endgroup

\end{document}